\documentclass[a4paper,11pt]{article}
\pdfoutput=1 

\usepackage{jheppub} 

\usepackage{gensymb}
\usepackage{subcaption}
\usepackage{soul}
\usepackage{slashed}
\usepackage{bm}
\usepackage{color}

\title{\boldmath Neutrino mass and leptogenesis in a hybrid seesaw model with a spontaneously broken CP}

\author{Rohan Pramanick,}
\author{Tirtha Sankar Ray}
\author{and Avirup Shaw}

\affiliation{Department of Physics, Indian Institute of Technology Kharagpur, Kharagpur 721302, India}

\emailAdd{rohanpramanick25@gmail.com}
\emailAdd{tirthasankar.ray@gmail.com}
\emailAdd{avirup.cu@gmail.com}

\abstract{We introduce a novel hybrid framework combining type I and type II seesaw models for neutrino mass where a complex vacuum expectation value of a singlet scalar field breaks CP spontaneously. Using pragmatic organizing symmetries we demonstrate that such a model can simultaneously explain the neutrino oscillation data and generate observed baryon asymmetry through leptogenesis. Interestingly, natural choice of parameters leads to a mixed leptogenesis scenario driven by nearly degenerate scalar triplet and right handed singlet neutrino fields for which we present a detailed quantitative analysis.}

\begin{document} 
\maketitle
\flushbottom

\section{Introduction} \label{sec:intro}
The lightness of the neutrino masses as indicated by  the oscillation data \cite{Fukuda:1998mi, Ahmad:2002jz, Ahn:2002up} remains as a stiking scar on the edifice of the Standard Model. A possible natural framework to address these issues is to invoke a seesaw Majorana mass for the neutrinos. This allows for the neutrino masses to be suppressed due to a large hierarchy in the weak scale and a neutrino scale set in the UV. Various realizations of the seesaw framework have been discussed in the literature \cite{deGouvea:2016qpx, Cai:2017jrq, Strumia:2006db, King:2003jb} including the more popular type I \cite{Minkowski:1977sc, Mohapatra:1979ia}, type II  \cite{Magg:1980ut, Schechter:1980gr, Cheng:1980qt, Lazarides:1980nt, Wetterich:1981bx, Mohapatra:1980yp} and the type III seesaw \cite{Foot:1988aq} mechanisms that can economically explain the smallness of neutrino mass.

Painstaking experimentation and  data analysis has progressively established the PMNS framework \cite{Pontecorvo:1957qd, Maki:1962mu} for the neutrino mass and mixing \cite{ParticleDataGroup:2022pth, Nunokawa:2007qh, Gonzalez-Garcia:2007dlo, Strumia:2006db}. In the three neutrino framework, the PMNS matrix has three mixing angles and one CP phase that determine the flavour oscillations. Additionally there are two Majorana phases which do not leave any imprint on the neutrino oscillation data. The mixing angles have been measured with increasing accuracy in various neutrino factories \cite{Giunti:2007ry}. For the more elusive CP violating phase $(\delta_{\rm CP})$ a non zero central value was implied at the T2K \cite{T2K:2019bcf} and NOvA \cite{NOvA:2019cyt}  experiment however with relatively large error bar for both normal hierarchy (NH) and inverted hierarchy (IH). This raised the tantalising possibility of having CP violation in the leptonic sector. Experimental observations limit the complex CKM phase in the quark sector to be insufficient to generate a sizable matter-antimatter asymmetry in the Universe \cite{ParticleDataGroup:2022pth}. The presence of a complex phase in the PMNS matrix that remains presently free from crippling experimental constraints, indicates CP violation in the neutrino sector, which is a necessary condition to generate observed baryon asymmetry of the universe (BAU) driven by leptogenesis \cite{Fukugita:1986hr}. The lepton asymmetry is generated by the CP violating out of equilibrium decay of heavy particles at the seesaw scale that can be converted into a baryon asymmetry by non perturbative sphaleron effects \cite{Khlebnikov:1988sr}.

In this work we present a scenario where the Standard Model (SM) is extended with an ${\rm SU(2)}_L$ triplet scalar with hypercharge $Y = 1$, an SM singlet Majorana fermion and a complex SM singlet scalar. The triplet scalar field and the right handed Majorana fermion generates a naturally light neutrino mass utilising the type II and type I seesaw mechanisms respectively. We present for the first time a minimal setup for spontaneously generated CP phase in the neutrino mass matrix driven by a hybrid type I and type II seesaw mechanism which complements the  existing literature \cite{Ferreira:2021bdj, Barreiros:2020gxu, Barreiros:2022aqu, Barreiros:2022fpi}. This model naturally incorporate  a novel degenerate leptogenesis framework where both the scalar triplet and the right handed neutrino simultaneously contribute to leptogenesis without any intervening washout phase. The scalar sector of the theory is constructed with a softly broken $Z_3$ symmetry that accommodates a CP violating complex vacuum expectation value (vev) for the singlet scalar field \cite{Lee:1973iz}. We demonstrate the propagation of the CP phase to the leptonic sector through the scalar triplet responsible for a type II like seesaw mass of the neutrinos which is complementary to the mechanism discussed in \cite {Branco:2003rt, Branco:2011zb}. The soft breaking that provides a solution to the domain wall problem is crucial for generating a nontrivial complex phase in the neutrino mass matrix. Keeping all Lagrangian parameters to be explicitly real and with the spontaneously broken CP we perform an extensive numerical scan of the model to identify the regions of parameter space that is in agreement with the global fit of  neutrino oscillation data \cite{Esteban:2020cvm}. Next we turn our attention to the possibility of generating the observed BAU \cite{Planck:2018vyg} driven by leptogenesis within this framework. Order one Yukawa couplings in the neutrino sector expectedly lead to a scenario of nearly degenerate triplet and right handed neutrino that simultaneously contribute to a mixed leptogenesis scenario without any intervening washout phase. We demonstrate that the model for spontaneous CP violation (SCPV)  presented here can simultaneously explain the neutrino mass and mixing while also generating a matter-antimatter asymmetry in consonance with observation.

This paper is organised as follows. In section \ref{sec: model} we present our model for neutrino masses combining the type I and type II seesaw mechanisms. In section \ref{sec: nu_osc_data} we present the results of our  parameter scans and discuss the fit to the global neutrino oscillation data. In section \ref{sec: lepto} we explore the possibility of leptogenesis within the framework of Singlet Doublet Triplet model before concluding in section \ref{sec: conclusion}.

\section{The Singlet Doublet Triplet model }\label{sec: model}

In this section we present a model which is a minimal extension of the SM augmented with one ${\rm SU(2)}_L$ singlet $(\sigma)$ and one complex triplet $(\Delta)$ with hypercharge $Y = 1$, in the fermion sector we introduce a singlet right handed Majorana neutrino $(N_R)$. We will refer to this model as the \emph{Singlet Doublet Triplet} (SDT) model in our discussions. The model provides an economical setup which accommodates the spontaneous violation of CP in the scalar sector and propagation of the CP phase in the neutrino sector where all the parameters are set to be real. The Majorana masses of the neutrinos are generated by a hybrid  seesaw mechanism that combined the type I and type II framework. While we follow a bottom up approach, for a UV completion of such hybrid mechanisms within SO(10) GUT or left-right symmetric models, see \cite{Akhmedov:2006de, Borah:2016iqd, Ohlsson:2019sja, Xing:2020ald}.

In the scalar sector we introduce a global discrete $Z_3$ symmetry that organises the minimal framework for spontaneous CP violation in the scalar sector. The relevant field content and their corresponding charges under ${\rm SU(2)}_L \times {\rm U(1)}_Y \times Z_3$ are summarised in Table \ref{tab: field_content} \footnote{We mention that imposition of $Z_8$ discrete symmetry group on the field contents leads towards similar results where the charge assignments on the fields have to be carried out accordingly.}.
\begin{table}[h!]
\centering
\renewcommand{\arraystretch}{1.25}
\begin{tabular}{|c|c|c|c|c|c|c|}
\hline 
\multicolumn{1}{|c|}{} & \multicolumn{3}{c|}{SM fields} & \multicolumn{3}{c|}{BSM fields}\tabularnewline
\hline 
 & $H$ & $L$ & $e_R$ & $\sigma$ & $\Delta$ & $N_R$\tabularnewline
\hline 
\hline 
${\rm SU(2)}_L$ & $2$ & $2$ & $1$ & $1$ & $3$ & $1$\tabularnewline
\hline 
$U(1)_Y$ & $\frac{1}{2}$ & $-\frac{1}{2}$ & $-1$ & $0$ & $1$ & $0$\tabularnewline
\hline 
$\mathbb{Z}_3$ & $1$ & $\omega$ & $\omega$ & $\omega$ & $\omega$ & $\omega$\tabularnewline
\hline 
\end{tabular}
\renewcommand{\arraystretch}{1.0}
\caption{Charge assignment and field content of the SDT model.}
\label{tab: field_content}
\end{table}

\subsection{Scalar potential}

The scalar potential for the model introduced in the previous section containing the field content given in Table \ref{tab: field_content} can be conveniently factorised in three parts
\begin{eqnarray}\label{eqn: tot_pot}
V(\sigma, H, \Delta) = V_\slashed{\text{CP}}(\sigma) + V_{II}(\sigma, H, \Delta) + V_\text{soft}(H, \Delta) \ ,
\end{eqnarray}
where $V_\slashed{\text{CP}}$ involving only the singlet field generates a complex vev of the singlet driving the SCPV.
\begin{eqnarray}
V_\slashed{\text{CP}}(\sigma) &=& -m_\sigma^2 \ \sigma^* \sigma + \lambda_\sigma \ (\sigma^* \sigma)^2 + \mu_{\sigma 3} \left[\sigma^3 + {\sigma^*}^3 \right]  \ . \label{eqn: pot_sigma_only}
\end{eqnarray}
The second part $(V_{II})$ generates an induced seesaw triplet vev while driving electroweak symmetry breaking as  
\begin{eqnarray}
V_{II}(\sigma, H, \Delta) &=& - \ m^2_H (H^\dag H)+\frac{\lambda }{4}(H^\dag H)^2 + m_\Delta^2 \text{Tr}(\Delta^\dag \Delta) + \lambda_2 \left[ \text{Tr}(\Delta^\dag \Delta)\right]^2  \nonumber \\
&& + \ \lambda_3 \text{Tr}(\Delta^\dag \Delta)^2 + \ \lambda_1 (H^\dag H) \text{Tr} (\Delta^\dag \Delta) + \lambda_4 (H^\dag \Delta \Delta^\dag H) \nonumber \\
&& + \ (\sigma^* \sigma) [\lambda_{\sigma H}(H^\dag H)  + \lambda_{\sigma \Delta} \text{Tr}(\Delta^\dag \Delta) ] \nonumber \\
&& + \ \lambda_{\sigma H \Delta} \left[ \sigma H^\mathsf{T} i \tau_2 \Delta^\dag H + \text{ h.c.} \right] \ , \label{eqn: V_all}
\end{eqnarray}
where the quartic coupling $\lambda_{\sigma H \Delta}$ is responsible for the propagation of the complex phase from the singlet vev to the triplet vev. The last term in the potential softly breaks the discrete $Z_3$ global symmetry given by
\begin{eqnarray}
V_\text{soft}(H, \Delta) &=& \mu \left[ H^\mathsf{T} i \tau_2 \Delta^\dag H + \text{ h.c.} \right] \ . \label{eqn: mu_term} 
\end{eqnarray}
The trilinear $\mu$ term not only prevents domain wall formation due to the  spontaneous breaking of discrete global symmetry but also is an essential requirement in order to generate CP phase in the neutrino mass matrix as we will see in the subsequent sections. In terms of their components the scalar fields are given as
\begin{eqnarray}\label{scalar_fields}
\sigma = \dfrac{1}{\sqrt{2}} \left( \sigma_R + i \sigma_I \right); \ H = \begin{pmatrix}
\phi^+ \\
\phi^0
\end{pmatrix}; \ \Delta = \begin{pmatrix}
\delta^+/\sqrt{2} & \delta^{++} \\
\delta^0		  & -\delta^+/\sqrt{2}
\end{pmatrix} \ ,
\end{eqnarray}
whereas the vev configurations of the corresponding fields are given by
\begin{eqnarray}\label{vev_config}
\left\langle \sigma \right\rangle = \dfrac{v_\sigma}{\sqrt{2}} e^{i \theta_\sigma}; \ H = \dfrac{1}{\sqrt{2}}\begin{pmatrix}
0 \\
v_H 
\end{pmatrix}; \ \Delta = \dfrac{1}{\sqrt{2}}\begin{pmatrix}
0 & \hspace{1em} 0 \\
v_\Delta e^{i \theta_\Delta} & \hspace{1em} 0
\end{pmatrix} \ .
\end{eqnarray}

\subsection{Yukawa sector and neutrino mass} \label{sec: yukawa_nu_order}

We now turn our attention to the neutrino mass matrix originating from the contribution of both type I and type II seesaw mechanisms. The Yukawa interactions consistent with the charge assignments of the fields given in Table \ref{tab: field_content} can be written as
\begin{eqnarray}
-\mathcal{L}_Y = \dfrac{1}{2} {\mathcal{Y}_\Delta}_{ij} L_i^\mathsf{T}\mathcal{C} i\tau_2 \Delta L_j + {\mathcal{Y}_\nu}_{i} \overline{L}_i \widetilde{H} N_R + \dfrac{1}{2} y_R \sigma^* \overline{N_R} N_R^c + {\mathcal{Y}_l}_{ij} \overline{L}_i H {e_R}_j + \text{ h.c.} \ .
\label{eqn: yukawa_lagrangian}
\end{eqnarray}
The Yukawa coupling matrices with the triplet, right handed neutrino (RHN) and the lepton doublet are denoted as $\mathcal{Y}_\Delta, \mathcal{Y}_\nu$ and $\mathcal{Y}_l$ respectively and $y_R$ is the Yukawa coupling responsible for generating the mass of the RHN after the singlet obtains a vev. We emphasize that all the parameters in Eq.~\ref{eqn: yukawa_lagrangian} are set to be real so that CP is explicitly conserved at the Lagrangian level. The triplet vev gives rise to a tiny neutrino mass in the type II seesaw mechanism as 
\begin{equation}
M_{II} = \dfrac{1}{\sqrt{2}}\mathcal{Y}_\Delta v_\Delta e^{i \theta_\Delta} \ .
\end{equation}

 However, an overall phase appearing in the complex mass matrix of the neutrinos can be reabsorbed in the field redefinition leading towards a vanishing CP phase $\delta_\text{CP}$ in the neutrino mass matrix, disfavoured by current experimental observations. This is easily fixed in the SDT model where the type I seesaw contribution triggered by the presence of bare Majorana mass term of the RHN introduces another in principle independent phase in the mass matrix 
\begin{eqnarray}
M_I = - \dfrac{1}{M_R} M_D M_D^\mathsf{T} = - \dfrac{\mathcal{Y}_\nu \mathcal{Y}_\nu^\mathsf{T}}{y_R} \dfrac{v_H^2}{2 v_\sigma} e^{i \theta_\sigma} \ ,
\end{eqnarray}
where $M_R = y_R v_\sigma e^{- i \theta_\sigma}$ is the complex Majorana mass of the RHN generated from the complex vev of the singlet field $\sigma$ and $M_D = \dfrac{1}{\sqrt{2}}\mathcal{Y}_\nu v_H$ is the Dirac mass of the active neutrinos generated from the vev of the usual SM Higgs doublet. The combined neutrino mass matrix is given by
\begin{equation}
\mathcal{L}_\text{mass} = - \dfrac{1}{2} \begin{pmatrix}
\overline{\nu^c_L} & \overline{N_R}
\end{pmatrix} M \begin{pmatrix}
{\nu_L} \\
N_R^c
\end{pmatrix}, \ \text{ where } 
M = \begin{pmatrix}
M_{II} & M_D \\
M_D^\mathsf{T} & M_R
\end{pmatrix} \ .
\end{equation}

With the field redefinitions as $N_R / L / e_R \rightarrow e^{i \theta_\sigma/2} (N_R / L / e_R)$ the neutrino masses can be approximately written as 
\begin{eqnarray}
M_\nu &\sim& | M_{II} | e^{i \theta_\text{eff}} - | M_{I} | + \mathcal{O} \left( \dfrac{1}{{|M_R|}^2} \right) \ , \\
M_\text{N} &\sim& |M_R| + \mathcal{O}\left(\dfrac{1}{|M_R|} \right) \ ,
\label{eqn: light_nu_mass_matrix}
\end{eqnarray}
where $\theta_\text{eff} = \theta_\Delta - \theta_\sigma$. This is indicative of the requirement for two independent contributions to the neutrino mass matrix with unequal phases. The explicit form of the light neutrino mass matrix is given as 
\begin{eqnarray} \label{eqn: explicit_nu_mass_matrix}
M_\nu = \dfrac{v_\Delta}{\sqrt{2}} \begin{pmatrix}
y_1 & y_2 & y_3 \\
y_2 & y_4 & y_5 \\
y_3 & y_5 & y_6 
\end{pmatrix} e^{i \theta_\text{eff} } - \dfrac{v_H^2}{2 |M_R|} \begin{pmatrix}
x_1^2 	& 	x_1 x_2 	& 	x_1 x_3 \\
x_1 x_2 &	x_2^2		& 	x_2 x_3	\\
x_1 x_3 &	x_2 x_3		&	x_3^2	
\end{pmatrix} \ ,
\end{eqnarray}
where $\mathcal{Y}_\nu = (x_1 \quad x_2 \quad x_3)^\mathsf{T}$ and we have redefined ${\mathcal{Y}_\Delta}_{11} = y_1$, ${\mathcal{Y}_\Delta}_{12} = y_2$, ${\mathcal{Y}_\Delta}_{13} = y_3$, ${\mathcal{Y}_\Delta}_{22} = y_4$, ${\mathcal{Y}_\Delta}_{23} = y_5 \text{ and } {\mathcal{Y}_\Delta}_{33} = y_6$ for simplicity. The neutrino mass matrix in our model is governed by nine Yukawa couplings, one mass $|M_R|$, a single phase $\theta_\text{eff}$ and the triplet vev $v_\Delta$. It is important to note that in the presence of the trilinear $\mu$ term, a non zero phase $\theta_\text{eff}$ gives rise to the CP violating phase $\delta_\text{CP}$ appearing in the neutrino mass matrix. 

The factorisation of the potential in Eq. \ref{eqn: tot_pot} is based on the separation of scales in the theory. This makes the discussion about the generation of the complex phase and its propagation to the neutrino mass matrix more tractable. Some comments about the justification of this factorisation is now in order. Simple back of the envelope estimates make these assertions clear. Given the measured neutrino masses are $\sim 10^{-11}$ GeV and assuming democratic contribution from the type I and type II seesaws we see that $v_\Delta \sim 10^{-11}$ GeV and $v_\sigma \sim 10^{15}$ GeV with 
\begin{eqnarray}
v_\Delta v_\sigma \sim v_H^2 \times \dfrac{\mathcal{O}(\mathcal{Y}_\nu)^2}{\mathcal{O}(\mathcal{Y}_\Delta)} \ .
\end{eqnarray}
This sets the scale associated with $V_\slashed{\text{CP}}$ to be at $10^{15}$ GeV that is well separated from the weak scale justifying using the decoupling limits to analyse the scalar potential. Given the single high scale set by the vev of the scalar singlet, both the solar and atmospheric neutrino mass gaps  are generated through an interplay of the Yukawa couplings as can be read off from Eq. \ref{eqn: explicit_nu_mass_matrix}.

\subsection{SCPV and generation of complex singlet vev}

The potential containing only the singlet $\sigma$ is given in Eq. \ref{eqn: pot_sigma_only} and consistent with the $Z_3$ discrete symmetry. The minimization conditions of potential are given as
\begin{eqnarray}
m_\sigma^2 - v_\sigma^2 \lambda_\sigma -\dfrac{3}{\sqrt{2}} v_\sigma \mu_{\sigma 3} \cos(3\theta_\sigma) &=& 0  \ , \\
\sin(3\theta_\sigma) &=& 0 \ ,
\end{eqnarray}
which implies a non zero complex vev for $\sigma$ given by
\begin{eqnarray}
m_\sigma^2 &=& v_\sigma^2 \lambda_\sigma - \dfrac{3}{\sqrt{2}} v_\sigma \mu_{\sigma 3} \ , \\
&& \theta_\sigma = \dfrac{\pi}{3} \ ,
\end{eqnarray}
and is responsible for the spontaneous breaking of the CP symmetry. The mass matrix in the basis $(\sigma_R \hspace{1em} \sigma_I)^\mathsf{T}$ is given as 
\begin{eqnarray}\label{Msigmasq}
\mathcal{M}_\sigma^2 = \begin{pmatrix}
\dfrac{1}{4} v_\sigma(v_\sigma \lambda_\sigma - 6 \sqrt{2} \mu_{\sigma 3}) \quad & \quad \dfrac{\sqrt{3}}{4} v_\sigma(v_\sigma \lambda_\sigma - 3 \sqrt{2} \mu_{\sigma 3}) \\
\dfrac{\sqrt{3}}{4} v_\sigma(v_\sigma \lambda_\sigma - 3 \sqrt{2} \mu_{\sigma 3}) \quad & \dfrac{3}{4} v_\sigma^2 \lambda_\sigma
\end{pmatrix} \ ,
\end{eqnarray}
which has a positive definite determinant given by 
\begin{eqnarray}
\text{Det} \mathcal{M}_\sigma = \dfrac{9}{8} v_\sigma^2 \mu_{\sigma 3} \left( 2 \sqrt{2} v_\sigma \lambda_\sigma - 3 \mu_{\sigma 3} \right) \ ,
\end{eqnarray}
upon choosing $\mu_{\sigma 3} >(<) \ 0 $ and $2 \sqrt{2} v_\sigma \lambda_\sigma > (<) \ 3 \mu_{\sigma 3}$. The mass eigenstates can be written as
\begin{eqnarray}
M_{\sigma 1}^2 &=& \dfrac{1}{\sqrt{2}} v_\sigma \left( 2 \sqrt{2} v_\sigma \lambda_\sigma - 3 \mu_{\sigma 3} \right)\;, \nonumber \\
&=& \dfrac{1}{4 \lambda_\sigma} \left[ 3 \mu_{\sigma 3} + \sqrt{9 \mu_{\sigma 3}^2 + 8 m_\sigma^2 \lambda_\sigma} \right] \sqrt{9 \mu_{\sigma 3}^2 + 8 m_\sigma^2 \lambda_\sigma}   \ , \\
M_{\sigma 2}^2 &=& \dfrac{9}{\sqrt{2}} v_\sigma \mu_{\sigma 3}\;, \nonumber \\
&=& \dfrac{9}{4 \lambda_\sigma} \left[ 3 \mu_{\sigma 3} + \sqrt{9 \mu_{\sigma 3}^2 + 8 m_\sigma^2 \lambda_\sigma} \right] \mu_{\sigma 3} \ .
\end{eqnarray}
It can be seen that in the limit $\mu_{\sigma 3} \rightarrow 0$, one of the mass eigenvalues $M_{\sigma 2}$ (which is mostly $\sigma_R$ like) goes to zero restoring the global $U(1)_\sigma$ symmetry of the potential and leads to CP conservation while the other eigenvalue $M_{\sigma 1}$ (which is mostly $\sigma_I$ like) remains positive definite.

\subsection{Generation of complex seesaw vev of the scalar triplet}
In this section we analyze the potential containing the Higgs doublet and the   triplet with the approximation that the singlet gets decoupled having a tiny mixing angles of $\mathcal{O}\left(v_{H, \Delta}/v_\sigma \right)$ with the doublet triplet sector. The potential containing the $H$ and $\Delta$ after the decoupling of the singlet can be written as
\begin{eqnarray}\label{V_del_H}
V_{II}(H, \Delta) &\approx& \left(-m^2_H + \dfrac{v_\sigma^2}{2} \lambda_{\sigma H} \right) H^\dag H + \frac{\lambda }{4}(H^\dag H)^2 + \left(m^2_\Delta + \dfrac{v_\sigma^2}{2} \lambda_{\sigma \Delta} \right) \text{Tr}(\Delta^\dag \Delta) \nonumber \\
&+& \lambda_2 \left[ \text{Tr}(\Delta^\dag \Delta)\right]^2 + \lambda_3 \text{Tr}(\Delta^\dag \Delta)^2 + \lambda_1 (H^\dag H) \text{Tr} (\Delta^\dag \Delta) + \lambda_4 (H^\dag \Delta \Delta^\dag H) \nonumber \\
&+& \left(\mu + \dfrac{v_\sigma}{\sqrt{2}} e^{i \theta_\sigma} \lambda_{\sigma H \Delta} \right) \left[ \sigma H^\mathsf{T} i \tau_2 \Delta^\dag H + \text{ h.c.} \right]  \ .
\end{eqnarray}
It is evident that the decoupling of the singlet leaves its imprint in the mass term of the doublet and the triplet and crucially in the effective trilinear term where we define 
\begin{equation}\label{eqn: mu_eff}
\widetilde{\mu} e^{i \widetilde{\theta}} = \mu + \dfrac{v_\sigma}{\sqrt{2}} e^{i \theta_\sigma} \lambda_{\sigma H \Delta} \ , 
\end{equation}
containing the only complex parameter in the potential. The minimization conditions are given as 
\begin{eqnarray}
- 4 m_H^2 + v_H^2 \lambda + 2 v_\Delta^2 \lambda_{14} + 2 v_\sigma^2 \lambda_{\sigma H} - 4\sqrt{2} v_\Delta \widetilde{\mu} \cos(\theta_\Delta - \widetilde{\theta}) &=& 0 \ , \label{mini1}\\
m_\Delta^2 + v_\Delta^2 \lambda_{23} + \dfrac{v_H^2}{2} \lambda_{14} + \dfrac{v_\sigma^2}{2} \lambda_{\sigma \Delta} - \dfrac{v_H^2 \widetilde{\mu}}{\sqrt{2} v_\Delta} \cos(\theta_\Delta - \widetilde{\theta}) &=& 0 \ , \label{mini2}\\
\sin ( \theta_\Delta - \widetilde{\theta}  ) &=& 0 \ , \label{mini3}
\end{eqnarray}
for $v_H, v_\Delta, \widetilde{\mu} \neq 0$ and $\lambda_{14} = \lambda_1 + \lambda_4$ and $\lambda_{23} = \lambda_2 + \lambda_3$. The masses and the phase of the triplet obtained from the solution of Eqs. \ref{mini1}, \ref{mini2} and \ref{mini3} are given by
\begin{eqnarray}
m_H^2 &=& \dfrac{1}{4} \left( v_H^2 \lambda + 2 v_\Delta^2 \lambda_{14} + 2 v_\sigma^2 \lambda_{\sigma H} - 4\sqrt{2} v_\Delta \widetilde{\mu} \right) \ , \label{mHsq}\\
m_\Delta^2 &=& - v_\Delta^2 \lambda_{23} - \dfrac{v_H^2}{2} \lambda_{14} - \dfrac{v_\sigma^2}{2} \lambda_{\sigma \Delta} + \dfrac{v_H^2  \widetilde{\mu}}{\sqrt{2} v_\Delta} \ , \label{mdelsq}\\
\theta_\Delta &=& \widetilde{\theta} = \tan^{-1} \dfrac{\lambda_{\sigma H \Delta} v_\sigma \sin \theta_\sigma}{\sqrt{2}\mu + \lambda_{\sigma H \Delta} v_\sigma \cos \theta_\sigma } \ . \label{eqn: th_delta}
\end{eqnarray}
It is important to note that the absence of the $Z_3$ breaking $\mu$ term in the potential leads to $\theta_\Delta = \theta_\sigma$ resulting in vanishing $\theta_\text{eff}$ in the neutrino mass matrix \footnote{It should be noted that the solution of $\theta_\Delta$ is arbitrary upto a additive $\pm \pi$.}.

We now move on to discuss the mass spectrum which is now more intricate in the absence of well defined CP charge. The real and imaginary parts of the neutral doublet and triplet field constitute a $4 \times 4$ mass matrix $\mathcal{M}_0^2$ given in the basis $(\phi^0_R \quad \delta^0_R \quad \phi^0_I \quad \delta^0_I)^\mathsf{T}$ where the fields are defined in Eq. \ref{scalar_fields} and the subscript R(I) refers to the Real (Imaginary) components, 
\begin{equation}
\mathcal{M}_0^2 = \begin{pmatrix}
\dfrac{1}{4} v_H^2 \lambda  \quad & \dfrac{1}{2}v_H \cos \widetilde{\theta} (v_\Delta \lambda_{14} - \sqrt{2} \widetilde{\mu}) \quad & 0 \quad & \dfrac{1}{2}v_H \sin \widetilde{\theta} (v_\Delta \lambda_{14} - \sqrt{2} \widetilde{\mu}) 
\\
\cdot & v_\Delta^2 \lambda_{23} \cos^2 \widetilde{\theta} + \dfrac{v_H^2 \widetilde{\mu} }{2\sqrt{2} v_\Delta} & \dfrac{1}{\sqrt{2}} v_H \widetilde{\mu} \sin \widetilde{\theta} & \dfrac{1}{2} v_\Delta^2 \lambda_{23} \sin 2 \widetilde{\theta} \\
\cdot & \cdot & \sqrt{2} v_\Delta \widetilde{\mu} & - \dfrac{1}{\sqrt{2}} v_H \widetilde{\mu} \cos \widetilde{\theta} \\
\cdot & \cdot & \cdot & v_\Delta^2 \lambda_{23} \sin^2 \widetilde{\theta} + \dfrac{v_H^2 \widetilde{\mu} }{2\sqrt{2} v_\Delta}
\end{pmatrix} \ \label{mass0sq},
\end{equation}
which can be diagonalized to get the mass eigenvalues. One of the eigenvalues is zero which corresponds to the Goldstone mode of the massive $Z$ boson, whereas the other three eigenvalues are given as
\begin{eqnarray}
M_{H_{1, 2}^0}^2 &\approx& \dfrac{1}{4 v_\Delta} \left[ v_H^2 (\sqrt{2} \widetilde{\mu} + \lambda v_\Delta) + 4 v_\Delta^3 \lambda_{23}  \mp \left\lbrace 16 v_\Delta^6 \lambda_{23}^2 + v_H^4 (v_\Delta^2 \lambda^2 - 2\sqrt{2} v_\Delta \lambda \widetilde{\mu} + 2 \widetilde{\mu}^2) \right. \right.  \nonumber \\ 
 && \left. \left. + \ 8 v_H^2 v_\Delta^2 \left( v_\Delta^2 (2 \lambda_1^2 - \lambda \lambda_{23} + 4 \lambda_1 \lambda_4 + 2 \lambda_4^2) + \sqrt{2} v_\Delta \widetilde{\mu} (\lambda_{23} - 4\lambda_{14}) + 4 \widetilde{\mu}^2 \right)  \right\rbrace^{1/2}  \right] \ , \label{mh12sq} \nonumber \\
M_{H_3^0}^2 &\approx& \dfrac{\widetilde{\mu}}{\sqrt{2} v_\Delta} \left( v_H^2 + 4 v_\Delta^2 \right) \ . \label{mh3sq}
\end{eqnarray}
Considering the order of magnitude estimation mentioned in section \ref{sec: yukawa_nu_order} it can be easily verified that $H_1^0$ corresponds to the usual $125$ GeV SM Higgs with mass $M_{H_1^0}^2 = \dfrac{\lambda}{4} v_H^2 +  \mathcal{O} \left(\dfrac{v_\Delta^2}{v_H^2} \right)$. On the other hand $H_2^0$ and $H_3^0$ corresponds to the real and imaginary part of the neutral triplet field with mass $\mathcal{O} (v_\sigma^2)$. It is also worth mentioning that in the absence of CP violation (i.e. $\widetilde{\theta} = 0$), the mass matrix $\mathcal{M}_0^2$ decomposes into two diagonal $2 \times 2$ blocks due to vanishing off diagonal elements. Similarly the singly charged scalar mass matrix in the basis $(\phi^+ \quad \delta^+)^\mathsf{T}$ is given as 
\begin{equation}
\mathcal{M}_+^2 = \left( \sqrt{2} \widetilde{\mu} - \dfrac{v_\Delta}{2} \lambda_4 \right) \begin{pmatrix}
v_\Delta \quad & -\dfrac{1}{\sqrt{2}} v_H \\
\cdot \quad & \dfrac{v_H^2}{2 v_\Delta}
\end{pmatrix} \ .
\end{equation}
The determinant of the matrix is zero which signifies one eigenvalue to be zero corresponding to the Goldstone mode of the $W$ boson. The masses of the singly and doubly charged mass eigenstates are given by
\begin{eqnarray}
M_{H^+}^2 &\approx& \dfrac{1}{4 v_\Delta} (v_H^2 + 2 v_\Delta^2) (2\sqrt{2} \widetilde{\mu} - v_\Delta^2 \lambda_4) \ , \label{mhpsq}\\
M_{H^{++}}^2 &\approx& \dfrac{v_H^2 \widetilde{\mu}}{\sqrt{2} v_\Delta} - \dfrac{1}{2} v_H^2 \lambda_4 - 2 v_\Delta^2 \lambda_3 \ . \label{mhpsq}
\end{eqnarray}
It is important to note that all the mass eigenstates of the triplet lies at the scale of the singlet vev of the order $10^{15}$ GeV. 

\section{Evaluation of neutrino oscillation parameters}\label{sec: nu_osc_data}

In order to extract the neutrino oscillation parameters we construct the matrix $h \equiv M_\nu M_\nu^\dagger$, where $M_\nu$ is the neutrino mass matrix from our model defined in Eq. \ref{eqn: light_nu_mass_matrix}. It can be diagonalized   by a unitary matrix $U$ in the following way
\begin{equation}
U^\dagger h U = \text{diag}(m_1^2, m_2^2, m_3^2) \ ,
\label{eqn: eigenval_eqn}
\end{equation}
where $m_1^2, m_2^2$ and $m_3^2$ are the squared eigenvalues of the $M_\nu$ matrix. Where the mixing matrix $U$ can be parameterized by three angles $\theta_{12}, \theta_{23}, \theta_{13}$ and one CP phase $\delta_\text{CP}$ following the PDG convention \cite{ParticleDataGroup:2020ssz} as 
\begin{equation}
U = \begin{pmatrix}
1 & 0 & 0 \\
0 & c_{23} & s_{23} \\
0 & -s_{23} & c_{23}
\end{pmatrix} \begin{pmatrix}
c_{13} & 0 & s_{13} e^{- i \delta_{\rm CP}} \\
0 & 1 & 0 \\
-s_{13} e^{i \delta_{\rm CP}} & 0 &  c_{13}
\end{pmatrix} \begin{pmatrix}
c_{12} & s_{12} & 0 \\
-s_{12} & c_{12} & 0 \\
0 & 0 & 1
\end{pmatrix} \ ,
\end{equation}
where $s_{ij} = \sin \theta_{ij}$, $c_{ij} = \cos \theta_{ij}$ are written for notational convenience. The explicit CP violation in the neutrino sector implies a non zero value for the rephasing invariant quantity defined as
\begin{equation}
J_{\rm CP} = \dfrac{\Im \big[ h_{12} h_{23} h_{31} \big]}{\Delta m_{21}^2 \Delta m_{31}^2 \Delta m_{32}^2 } \ ,
\label{eqn: J_in_h}
\end{equation}
where $\Delta m_{ij}^2 = m_i^2 - m_j^2$. Within the three active neutrino framework, the $J_\text{CP}$ can  be expressed in terms of three mixing angles and one Dirac CP phase,
\begin{equation}
J_\text{CP} = \dfrac{1}{8} \sin(2\theta_{12}) \sin(2\theta_{23}) \sin(2\theta_{13}) \cos\theta_{13} \sin\delta_{\rm CP} \ .
\label{eqn: J_in_del}
\end{equation}
The $h_{12}h_{23}h_{31}$ appearing in  Eq. \ref{eqn: J_in_h} can be expressed in terms of model parameters correlating the CP phase $\delta_{\rm CP} $ with the spontaneously generated $\theta_{\rm eff}.$ Analytic expressions are  given in Appendix \ref{ap4}.

We use the six global fit parameters (two mass squared differences, three angles and one phase) extracted from neutrino oscillation data \cite{Esteban:2020cvm} given in Table \ref{tab: nu_osc_data} to put constraints on the parameter space of our model.
\begin{table}[h]
\begin{center}
\resizebox{1\textwidth}{!}{%
\renewcommand{\arraystretch}{1.5}
\begin{tabular}{|c|c||c|c|}
\hline 
Parameter & NH (IH) & Parameter & NH (IH) \tabularnewline
\hline 
\hline 
$\Delta m_{21}^2 \big/ 10^{-5} \text{ eV}^2$ & $ 6.82 - 8.04 $ & $\theta_{12}$ & $ 31.27 \degree - 35.86 \degree (35.87 \degree) $  \tabularnewline
\hline 
$\Delta m_{31}^2 \big/ 10^{-3} \text{ eV}^2$ & $ 2.431(-2.581) - 2.598(-2.414) $ & $\theta_{23}$ & $ 40.1\degree (40.3 \degree) - 51.7\degree (51.8\degree) $ \tabularnewline
\hline 
$\delta_\text{CP}$ & $ 107\degree (193\degree) - 403\degree (352\degree) $  & $\theta_{13}$ & $ 8.20\degree (8.24\degree) - 8.93\degree (8.96 \degree) $  \tabularnewline
\hline 
\end{tabular}
\renewcommand{\arraystretch}{1.0}
}
\end{center}
\caption{$3 \sigma$ intervals for the neutrino masses and mixing parameters from global fits of experimental neutrino oscillation data \cite{Esteban:2020cvm}. }
\label{tab: nu_osc_data}
\end{table}

\begin{table}[b]
\centering
\begin{center}
\renewcommand{\arraystretch}{1.25}
\begin{tabular}{|c|c||c|c|}
\hline 
Parameter & Range & Parameter & Range\tabularnewline
\hline 
\hline 
$y_i$ where $i = 1, 2, \cdots 6$ & $\big[10^{-3}, \ 10^1 \big]$ & $x_i$ where $i = 1, 2, 3$ & $\big[10^{-3}, \ 10^1 \big]$\tabularnewline
\hline
$\lambda_{\sigma H \Delta}$ & $\big[10^{-3}, \ 10^1 \big]$ & $ \mu$ & $\big[10^{9}, \ 10^{12} \big]$ GeV\tabularnewline
\hline 
$|M_R|$ & $10^{15}$ GeV & $v_\Delta$ & $10^{-11}$ GeV \tabularnewline
\hline 
\end{tabular}
\renewcommand{\arraystretch}{1.0}
\end{center}
\caption{Ranges of free parameters used in the SDT model.}
\label{tab: param_range}
\end{table}

The neutrino mass matrix defined in terms of the model parameters given in Eq. \ref{eqn: light_nu_mass_matrix} consists of nine real Yukawa-like couplings and one effective phase $\theta_\text{eff}$ which is a function of underlying model parameters, the trilinear coupling $\mu$ and the quartic coupling $\lambda_{\sigma H \Delta}$. The other parameters are the triplet vev $v_\Delta$ and the mass of the RHN $|M_R|$ which are set to $10^{-11}$ GeV and $10^{15}$ GeV respectively as discussed in section \ref{sec: yukawa_nu_order}. We have varied the free parameters of our model within the ranges listed in Table \ref{tab: param_range} to reproduce the neutrino oscillation data for NH and IH. The ranges of Yukawa-like couplings ($y$'s and $x$'s) and the quartic coupling $(\lambda_{\sigma H \Delta})$ are within the perturbative limit in our analysis. The range of the trilinear coupling $(\mu)$ is set while ensuring the absence of tachyonic modes \cite{Arhrib:2011uy}. We construct the neutrino mass matrix and extract various neutrino oscillation parameters following the algorithm described in \cite{Adhikary:2013bma} and compare with various experimental constraints listed below.
\begin{enumerate}
\item The oscillation parameters given in Table \ref{tab: nu_osc_data} adapted from the NuFIT analysis \cite{Esteban:2020cvm} at $3 \sigma$.
\item We set the cosmological upper limit on the sum over neutrino masses $\sum m_\nu < 0.12$ eV from the latest Planck data \cite{Planck:2018vyg}.

\item The best fit value of the Jarlskog invariant is set at $J_\text{CP}^\text{best} = - 0.019$ and considered $3\sigma$ range to obtain allowed parameter points \cite{ParticleDataGroup:2020ssz}.

\item We constraint the effective Majorana neutrino parameter $|m_{\beta \beta}|$ \cite{Dolinski:2019nrj} from the relevant experiments including (i) GERDA II with $m_{\beta \beta} < (0.079 - 0.180)$ eV \cite{GERDA:2020xhi}, (ii) CUORE-0 with $m_{\beta \beta} < (0.110 - 0.520)$ eV \cite{CUORE:2017tlq} and (iii) KamLAND-Zen Collaboration with $m_{\beta \beta} < (0.061 - 0.165)$ eV \cite{KamLAND-Zen:2016pfg}.
\end{enumerate}
We have performed an extensive eleven dimensional numerical scan over the nine free Yukawa parameters and two parameters of scalar potential utilising a dedicated Markhov Chain Monte Carlo algorithm to find out the parameter space in agreement with the above mentioned constraints. Out of $10^{10}$ parameter points $\sim 7 \times 10^5 \ (0.007 \%)$ and $\sim 2.5 \times 10^5 \ (0.0025 \%)$ points satisfy with NH and IH of neutrino mass respectively. 

\begin{figure}[t]
\centering
\begin{subfigure}{0.49\textwidth}
\centering
\includegraphics[width=1\linewidth]{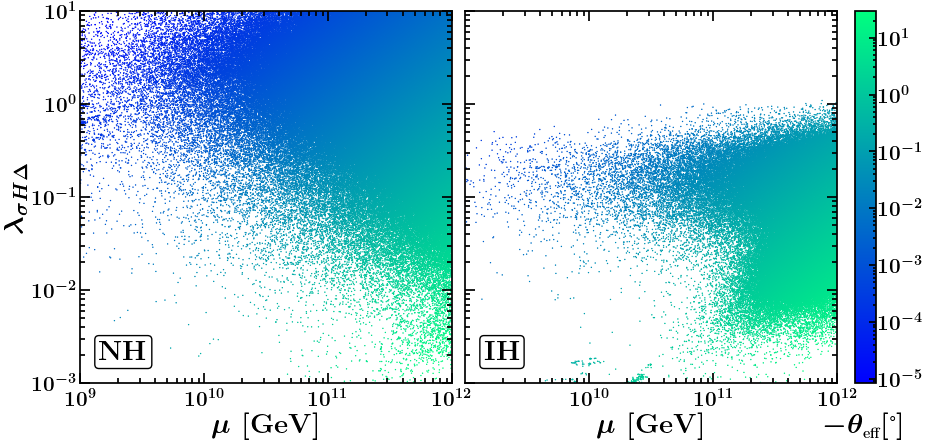} 
\caption{}
\label{fig1: mu_lam_theff_NH_IH}
\end{subfigure}
\begin{subfigure}{0.49\textwidth}
\centering
\includegraphics[width=1\linewidth]{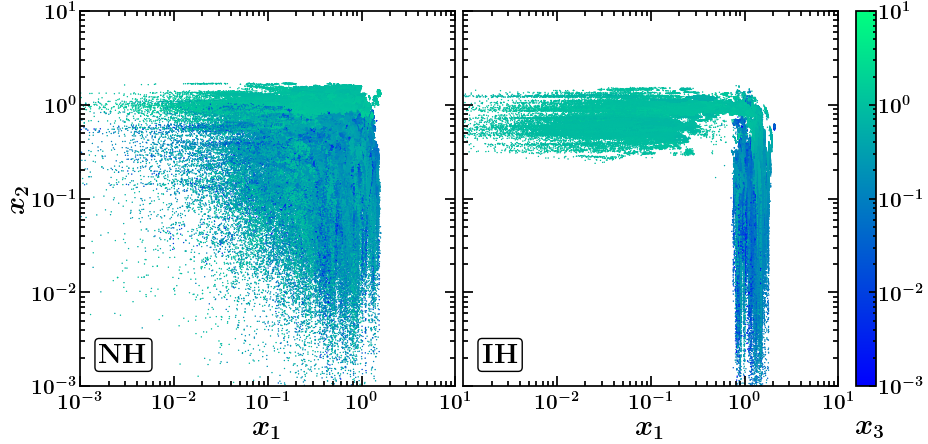} 
\caption{}
\label{fig1: x1_x2_x3_NH_IH}
\end{subfigure}
\vspace{2.5ex} \\
\begin{subfigure}{0.49\textwidth}
\centering
\includegraphics[width=1\linewidth]{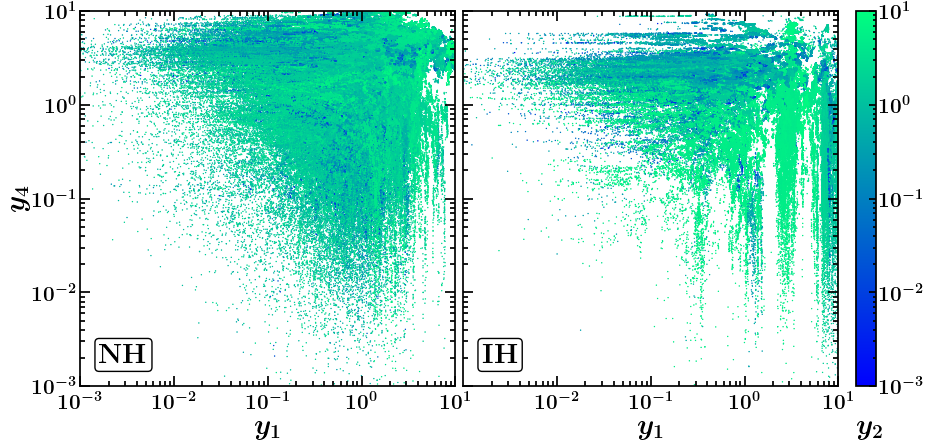} 
\caption{}
\label{fig1: y1_y4_y2_NH_IH}
\end{subfigure}
\begin{subfigure}{0.49\textwidth}
\centering
\includegraphics[width=1\linewidth]{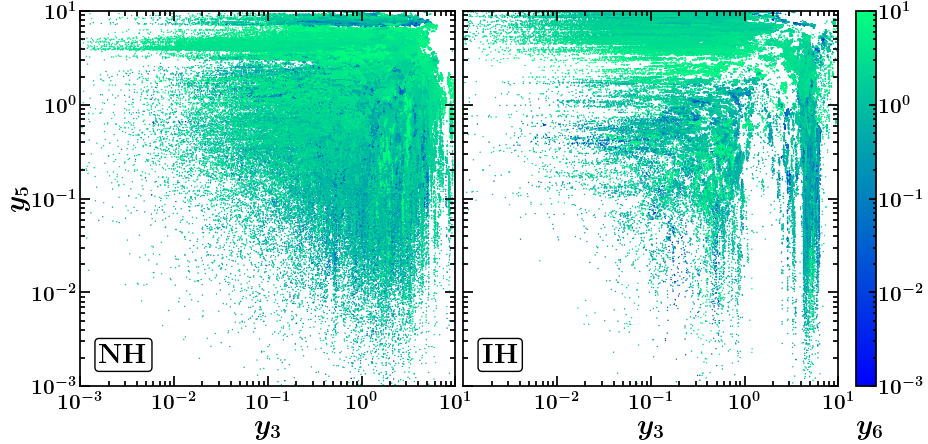} 
\caption{}
\label{fig1: y3_y5_y6_NH_IH}
\end{subfigure}
\caption{Model parameter space satisfying neutrino oscillation data for both normal hierarchy (NH) and inverted hierarchy (IH).} 
\label{fig: param_space_nu}
\end{figure}

The model parameter space is shown in Fig. \ref{fig: param_space_nu}. The value of $\theta_\text{eff}$ decreases as the deviation from $\theta_\sigma$ increases due to increase in trilinear coupling which is evident from Fig. \ref{fig1: mu_lam_theff_NH_IH} while being consistent with the minimization condition of $\theta_\Delta$ in Eq. \ref{eqn: th_delta}.  The Yukawa-like couplings of most of the points are concentrated about unity which is also in harmony with the order of magnitude estimation carried out in section \ref{sec: yukawa_nu_order} as is evident from Figs. \ref{fig1: x1_x2_x3_NH_IH}, \ref{fig1: y1_y4_y2_NH_IH} and \ref{fig1: y3_y5_y6_NH_IH}.
\begin{figure}[t]
\centering
\begin{subfigure}{0.485\textwidth}
\centering
\includegraphics[width=1\linewidth]{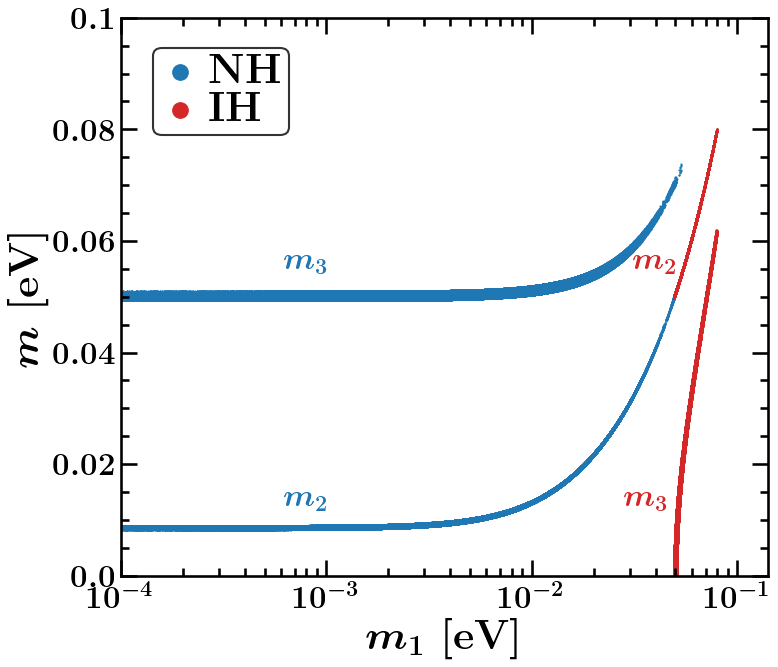}
\caption{}
\label{fig: m1_m2_m3} 
\end{subfigure}
\begin{subfigure}{0.505\textwidth} 
\centering
\includegraphics[width=1\linewidth]{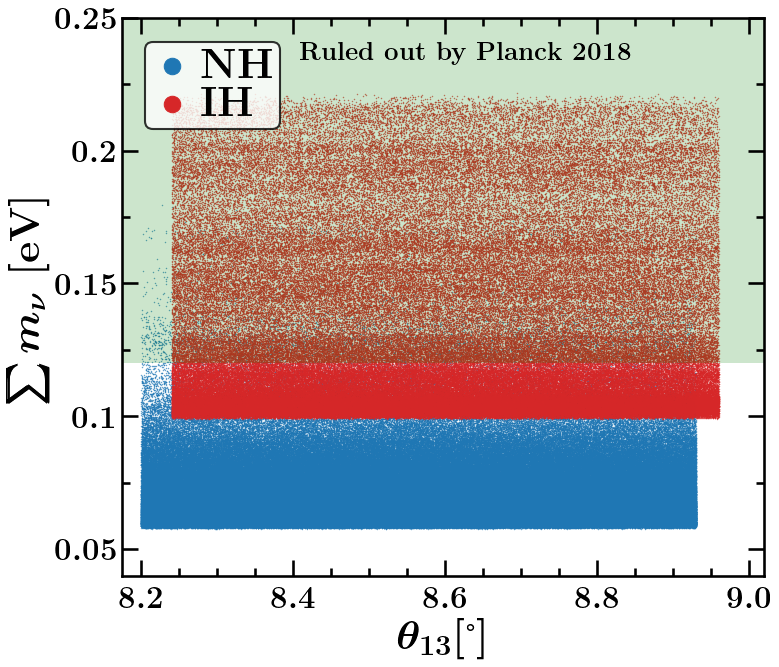} 
\caption{}
\label{fig: th_totalmass}
\end{subfigure}
\caption{Left panel (\ref{fig: m1_m2_m3}): Absolute mass of neutrinos allowed by neutrino oscillation data. Right panel (\ref{fig: th_totalmass}): Total mass of active neutrinos with respect to the reactor mixing angle $\theta_{13}$. The blue (red) points are allowed by neutrino oscillation data for NH (IH) and the shaded green region is excluded from Planck data.} 
\end{figure}
\begin{figure}[t]
\centering
\begin{subfigure}{0.49\textwidth}
\centering
\includegraphics[width=1\linewidth]{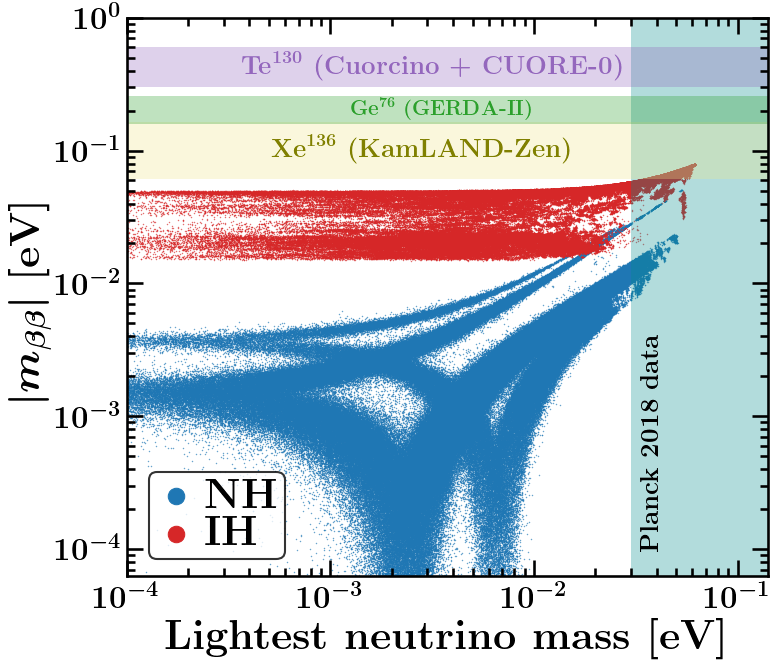} 
\caption{}
\label{fig: majorana_mass_plot}
\end{subfigure}
\begin{subfigure}{0.49\textwidth}
\centering
\includegraphics[width=1\linewidth]{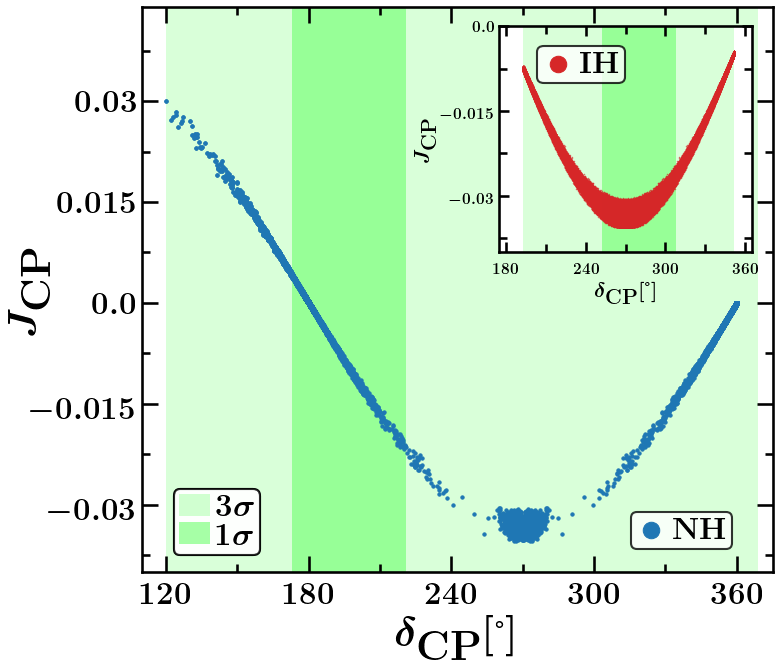} 
\caption{}
\label{fig: delta_jcp_plot}
\end{subfigure}
\caption{Left panel (\ref{fig: majorana_mass_plot}): Effective Majorana mass variation with respect to the lightest neutrino mass. Right panel (\ref{fig: delta_jcp_plot}): Jarlskog invariant as a function of the CP violating phase. The green shaded region indicates different confidence intervals in the measurement of $\delta_{\rm CP}$. The blue (red) points are allowed by neutrino oscillation data for NH (IH). }
\end{figure}

The absolute mass of the neutrinos allowed by the neutrino oscillation data \cite{Esteban:2020cvm} is shown in Fig. \ref{fig: m1_m2_m3}. Expectedly the small solar mass gap $\Delta m_{21}^2 \sim \mathcal{O}(10^{-5} \text{ eV}^2)$ results in the allowed parameter space to shrink considerably in the $m_1 - m_2$ plane, resembling $m_1 \simeq m_2$ for both normal and inverted mass ordering. Given the larger atmospheric mass gap $\Delta m_{31}^2 \sim \mathcal{O}(10^{-3} \text{ eV}^2)$ the $m_3$ values are above (below) the $m_1 - m_2$ line given that $m_3 \gg m_1 (m_1 \gg m_3)$ for NH (IH) scenario. In Fig. \ref{fig: th_totalmass} the total mass of all the active neutrinos $\sum m_\nu$ is depicted with respect to the reactor mixing angle $\theta_{13}$. There is a cosmological upper bound on the sum over neutrino masses as measured by the Planck Collaboration \cite{Planck:2018vyg}. As a result a considerable portion of allowed points for IH gets disfavoured while almost the entire region of NH remains consistent.

The effective Majorana mass parameter defined as $m_{\beta \beta} = \big\vert \sum_i m_i U_{ei}^2 \big \vert $ that is directly sensitive to neutrinoless double beta decay experiments have been plotted against the lowest neutrino mass in Fig. \ref{fig: majorana_mass_plot}. Similarly the allowed region in the parameter space of the Jarlskog invariant $J_\text{CP}$ and the CP violating phase $\delta_\text{CP}$ is displayed in Fig. \ref{fig: delta_jcp_plot}. In our numerical scan we have taken care of the degeneracies arising from the inverse of a sine function in estimating $\delta_\text{CP}$. As can be read off from these two figures a large number of data points remains in agreement with the neutrinoless double beta decay experiments and measurements of the CP phase at neutrino factories. 

We conclude that the SDT model is capable of surviving the present neutrino constraints. Armed with this we will now explore the possibility of  leptogenesis within the framework of SDT model.

\section{Leptogenesis within the SDT framework}\label{sec: lepto}

The observed baryon asymmetry of the Universe as extracted from the studies of the Cosmic Ray Microwave Background at present stand at \cite{Planck:2018vyg} 
\begin{eqnarray}\label{bau}
&&Y_B=8.750\pm 0.077 \times 10^{-11} \ .\label{YB}
\end{eqnarray}
Among the several possible mechanisms to study baryogenesis such as GUT baryogenesis \cite{Ignatiev:1978uf,Yoshimura:1978ex,Toussaint:1978br,Dimopoulos:1978kv,Ellis:1978xg,Weinberg:1979bt,Yoshimura:1979gy,Barr:1979ye,Nanopoulos:1979gx,Yildiz:1979gx}, Affleck-Dine mechanism \cite{Affleck:1984fy,Dine:1995kz}, electroweak baryogenesis \cite{Rubakov:1996vz,Riotto:1999yt,Cline:2006ts}, baryogenesis via leptogenesis \cite{Fukugita:1986hr,Riotto:1999yt,Pilaftsis:1997jf,Buchmuller:2004nz,Buchmuller:2005eh,Abada:2006ea,Davidson:2008bu,DiBari:2015oca} is very popular. In the following we are going to measure baryon asymmetry using leptogenesis in the present framework.

From the interaction terms given in Eqs.~\ref{eqn: V_all}, \ref{eqn: mu_term} and \ref{eqn: yukawa_lagrangian}, it is evident that there is a possibility of violation of lepton number due to the concurrence of a complex phase in the Yukawa interaction from the type I seesaw interaction and type II seesaw interaction. Within the SDT framework there are two different sources of CP asymmetry: (i) decay of right handed neutrino ($N_R$) to lepton and Higgs pair and (ii) decay of scalar triplet ($\Delta$) to lepton pair. We now systematically discuss the mixed leptogenesis framework from the SDT model.

\subsection{CP asymmetry parameter}

For CP violating decay of the $N_R$ to leptons and Higgs pair the CP asymmetry is given as follows
\begin{equation}
\epsilon_{N}=\sum_i {{\Gamma (N \rightarrow l_i + H^*) - 
\Gamma (N \rightarrow 
\bar{l}_i + H)}\over{\Gamma (N \rightarrow l_i + H^*) +
\Gamma (N \rightarrow 
\bar{l}_i + H)}} \,.
\end{equation} 

\begin{figure}[htbp!]
\centering
\subfloat[]{\label{NR}\includegraphics[scale=0.6,angle=0]{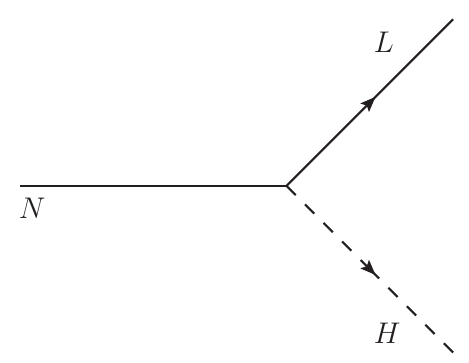}}
\subfloat[]{\label{NRloop}\includegraphics[scale=0.5,angle=0]{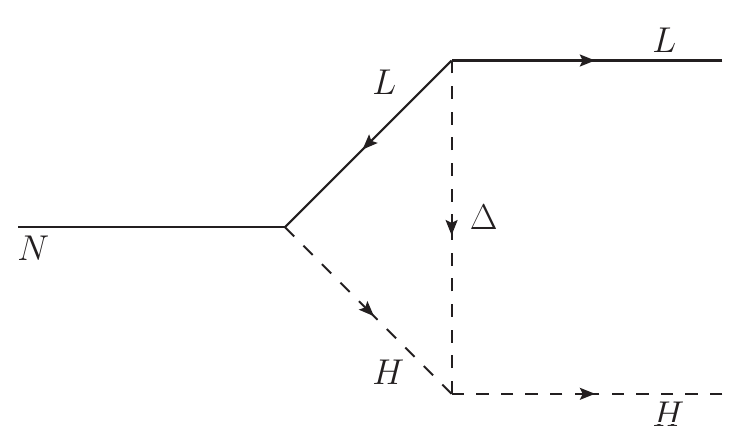}}\\
\caption{Tree level (left panel) and triplet mediated one loop (right panel) diagrams giving rise to CP asymmetry from the $N_R$ decay.}
\label{fig:N}
\end{figure}

In the present scenario this asymmetry is obtained from the interference of the tree level (Fig. \ref{NR}) with the one loop induced decay of $N_R$ involving a virtual $\Delta$ (Fig. \ref{NRloop}) and evaluates to \cite{Hambye:2003ka}
\begin{equation}
\epsilon_{N}^\Delta=-\frac{1}{16 \pi^2 \Gamma_N} \sum_{il} 
{\cal I}m[\mathcal{Y}_{\nu i} 
\mathcal{Y}_{\nu l}(\mathcal{Y}_{\Delta il} e^{i \theta_\text{eff}})^* \widetilde{\mu}^\ast] \, 
\Big(1-\frac{M^2_\Delta}{|M_R|^2} \log(1+|M_R|^2/M^2_\Delta) \Big) \ ,
\label{epsN}
\end{equation}
where $\Gamma_N$ is the tree level decay width of $N_R$ and is given by \cite{Hambye:2003ka}
\begin{equation}
\Gamma_{N}= \frac{1}{8 \pi} |M_R| \sum_i |(\mathcal{Y}_{\nu i})|^2 \,.
\label{gammaN}
\end{equation}

\begin{figure}[h!]
\centering
\subfloat[]{\label{Del}\includegraphics[scale=0.6,angle=0]{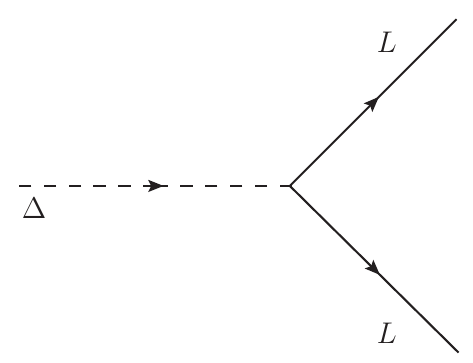}}
\subfloat[]{\label{Delloop}\includegraphics[scale=0.5,angle=0]{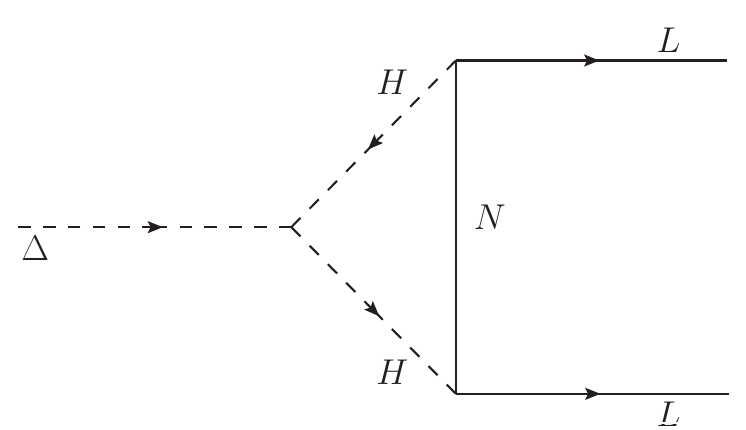}}
\caption{Tree level (left panel) and RHN mediated one loop (right panel) diagrams contributing to CP asymmetry from the $\Delta$ decay.}
\label{fig:Del}
\end{figure}

Within the present framework CP asymmetry can also originate from the decay of the scalar triplet $\Delta$ to same sign di-leptons. Similar to the previous case this originates from the tree level (Fig. \ref{Del}) and the one loop diagram (Fig. \ref{Delloop})  involving a virtual $N_R$ and is  given by \cite{Hambye:2003ka},
\begin{eqnarray}
\epsilon_\Delta^N &=& 2\cdot
{{\Gamma (\Delta^* \rightarrow l + l) - \Gamma (\Delta \rightarrow 
\bar{l} + \bar{l })}\over{\Gamma (\Delta^* \rightarrow l + l) 
    +\Gamma (\Delta \rightarrow \bar{l} + \bar{l}     )}} \ , \\
&=& \frac{|M_R|}{64 \pi^2 M_\Delta \Gamma_\Delta} \sum_{il}{\cal I}m[\mathcal{Y}^*_{\nu i} 
\mathcal{Y}^*_{\nu l}(\mathcal{Y}_{\Delta il} e^{i \theta_\text{eff}}) \widetilde{\mu} ]
\log(1+M^2_\Delta/|M_R|^2) \,,
\label{epsD}
\end{eqnarray}
where $\Gamma_\Delta$ is the total decay (contribution to two leptons and two scalars) width of triplet given as \cite{Hambye:2003ka}
\begin{equation}
\Gamma_\Delta=\frac{1}{8 \pi} M_\Delta \Big( \sum_{ij}|(\mathcal{Y}_{\Delta ij})|^2 +
\frac{\widetilde{\mu}^2}{M^2_\Delta} \Big) \,.
\label{gammaD}
\end{equation}
It is clearly evident from the Eqs. \ref{epsN} and \ref{epsD} that both the CP asymmetry parameters are proportional to $\widetilde{\mu}$ which captures the relative phase between the type I and type II neutrino mass terms. This single phase which generates CP asymmetry in the PMNS neutrino matrix can drive leptogenesis in the SDT model.
 
\subsection{\boldmath$B-L$ evolution} 
The lepton asymmetry originated via leptogenesis at the seesaw scale, can be connected to the present day value by solving  Boltzmann equations (BEs) which govern the out-of-equilibrium dynamics of RHN and scalar triplet involving processes (particularly in our case). Our goal is to identify only those reactions within the hot plasma that have decay rates comparable to Hubble rate at that temperature, i.e., $\Gamma (T) \sim H(T)$. In the passing we note that the study of leptogenesis with RHN and scalar triplet field have been discussed in several references \cite{Hambye:2003ka, Gu:2006wj, Chan:2007ng, AristizabalSierra:2011ab,Sierra:2014tqa, Chakraborty:2019uxk, Datta:2021gyi}.

The interaction terms given in Eqs.~\ref{eqn: V_all}, \ref{eqn: mu_term} and \ref{eqn: yukawa_lagrangian} violate lepton number by one or two units whenever $N_R$ decays to $(l,H)$ pair or $\Delta$ decays to $(l_i,l_j)$, respectively keeping the baryon number conserved. Our objective is to investigate the evolution of $(B-L)$ abundance with the understanding that the lepton asymmetry is translated to a baryon asymmetry through SM interaction possibly through the non perturbative sphaleron processes \cite{Khlebnikov:1988sr}.

Neutrino oscillation data assuming $\mathcal{O}(1)$ Yukawa couplings, imply that the masses of both $N_R$ and $\Delta$ are $\sim 10^{15}$ GeV with high degree of degeneracy. It should be noted that at this temperature none of the quark or lepton Yukawas are in thermal equilibrium suppressing generation of flavour asymmetry. Additionally due to degenerate mass and similar Yukawa couplings of $N_R$ and $\Delta$, the generation of flavour effects through washout is numerically insignificant \cite{Engelhard:2006yg}. With this caveat, the relevant BEs leading to unflavoured leptogenesis \cite{AristizabalSierra:2011ab,Sierra:2014tqa} are given by,
\begin{eqnarray}
&& { \dot{Y}_{N}=-\Big(\frac{Y_{N}}{Y_{N}^{eq}}-1 \Big ) \gamma_{D_{N}}}\;,\label{boltz_uf1}\\
&& \dot{Y}_\Sigma=-\Big(\frac{Y_\Sigma}{Y_\Sigma^{eq}}-1 \Big )\gamma_D -2\Big[\Big(\frac{Y_\Sigma}{Y_\Sigma^{eq}}\Big )^2-1 \Big]\gamma_A\;, \label{boltz_uf2}\\ 
&& \dot{Y}_{\Delta_\Delta}=-\Big[\frac{Y_{\Delta_\Delta}}{Y_\Sigma^{eq}}-\sum_{k} \Big( B_l C_k^l -B_H C_k^H \Big)\frac{Y_{\Delta_k}}{Y_l^{eq}}\Big]\gamma_D\;,\label{boltz_uf3}\\ 
&& \dot{Y}_{\Delta_{B-L}}= {-\Big[\Big(\frac{Y_{N}}{Y_{N}^{eq}}-1 \Big ) \varepsilon_{N}^{\Delta}+\Big( \sum_k C^l_{k}\frac{ Y_{\Delta_k}}{Y_l^{eq}} +
\sum_k C^H_k \frac{Y_{\Delta_k}}{Y_l^{eq}} \Big)\Big] \gamma_{D_{N}} }\nonumber\\
&&\hspace{1.7cm}-\Big[ \Big( \frac{Y_\Sigma}{Y_\Sigma^{eq}} -1 \big) \varepsilon_{\Delta}^N -2 \Big( \frac{Y_{\Delta_\Delta}}{Y_\Sigma^{eq}} -
 \sum_k C^l_{k} \frac{Y_{\Delta_k}}{Y_l^{eq}}\Big)B_{l} \Big]\gamma_D \nonumber\\
&&\hspace{1.7cm}-2 \sum_k \Big( C^H_k +C^l_k \Big ) \frac{Y_{\Delta_k}}{Y^{eq}_l} \Big (\gamma^{\prime HH}_{ll} +
\gamma^{H l}_{H l} \Big )\ ,
 \label{boltz_uf4}
\end{eqnarray}
where symbols have usual meaning. Here $n_X~(n_{\bar{X}})$ is the number density of $X~({\bar X})$  and the corresponding expressions for various particles are given in Appendix \ref{sec: appendix_a}. All the variables within the differential equations $(Y_N, Y_{\Delta_\Delta},Y_\Sigma,Y_{\Delta_{ B-L}})$ are functions of $z=M_\Delta/T$ and $\dot{Y}_X\equiv\dot{Y}_X(z)=zs(z)H(z)\frac{dY_X(z)}{dz}$. The scalar triplet density and asymmetry are defined as $\Sigma=\Delta+\Delta^\dagger$ and $\Delta_\Delta=\Delta-\Delta^\dagger$, respectively. The superscript \textquoteleft$eq$\textquoteright~ stands for the equilibrium values of the relevant quantities. In the Appendix \ref{sec: appendix_a} we list the the closed form expressions of all such equilibrium densities. Further, $B_l$ and $B_H$ represent branching ratios of $\Delta$ decaying to leptons and $HH$, respectively and their expressions are given as follows
\cite{Sierra:2014tqa} 
\begin{eqnarray}
&& B_l=\sum_{i,j=e,\mu,\tau} \frac{M_\Delta}{8\pi \Gamma_\Delta} |(\mathcal{Y}_{\Delta_{ij}} )|^2\;,\\
&& B_H =\frac{|\mu^{\rm eff}|^2}{8\pi M_\Delta \Gamma_\Delta}\;. 
\end{eqnarray}
Here, $\Gamma_\Delta$ is the total decay width of $\Delta$, while $B_l+B_H=1$. In the BEs, the quantity $\gamma_D$ ($\gamma_{D_N}$) represents the total reaction density of the $\Delta ~(N)$ incorporating its decay and inverse decay to Higgs pair and lepton pair (lepton and Higss pair). $\gamma_A$ denotes the reaction density for gauge boson mediated $2\leftrightarrow2$ scattering of triplet to gauge bosons, scalars and fermions. Further, $\gamma^{H H}_{l_i l_j}$ and $\gamma^{H l_j}_{H l_i}$ represent the reaction densities regarding lepton number $(\Delta L=2)$ and flavour violating Yukawa scalar mediated s-channel $(H H \leftrightarrow \bar{l_i} \bar{l_j})$ and $t-$ channel $(H \bar{l_i} \leftrightarrow H \bar{l_j})$ scattering respectively. The primed s-channel reaction densities are defined by $\gamma^\prime=\gamma-\gamma^{\rm (on~ shell)}$.  Here $C^l$ associate the asymmetry of lepton doublets with that of $B-L$, while $C^H$ relates the  the asymmetry of scalar triplet and $B-L$, 
e.g., $Y_{\Delta_{l}}=-\sum_k C^l_{k} Y_{\Delta_k}$\;,~~$Y_{\Delta_H}=-\sum_k C^H _k Y_{\Delta_k}$, where $Y_{\Delta_k}$ is the $k^{\rm th}$ component of the asymmetry vector $\vec{Y}_{\Delta}$ which can be represented as
\begin{equation}
\vec{Y}_{\Delta} \equiv  (Y_{\Delta_\Delta}, Y_{\Delta_{B-L}} )^T .
\end{equation}
The matrices $C^l$ and $C^H$ are obtained from different chemical equilibrium conditions. The corresponding values (applicable in our case) are taken from \cite{Sierra:2014tqa}.

Finally, simultaneous solutions of four differential equations (Eqs. \ref{boltz_uf1}, \ref{boltz_uf2}, \ref{boltz_uf3}  and \ref{boltz_uf4}) give us the value of all the asymmetry parameters. The abundance of $(Y_{\Delta_{B-L}})$ saturates at large value of $z$. Eventually the baryon asymmetry can be obtained via sphaleron processes as given by \cite{AristizabalSierra:2011ab,Sierra:2014tqa}
\begin{equation}
Y_{\Delta_B}=3 \times \frac{12}{37}\sum_i Y_{\Delta_{B-L}}\;, \label{yb}
\end{equation}
where the factor $3$ indicates the degrees of freedom of $\Delta$. 

\subsection{The SDT model parameter space} \label{sec: lepto_param_scan}

\begin{figure}[h!]
\begin{center}
\hspace*{-1cm}
\subfloat[]{\label{mu-lambda}\includegraphics[scale=0.175,angle=0]{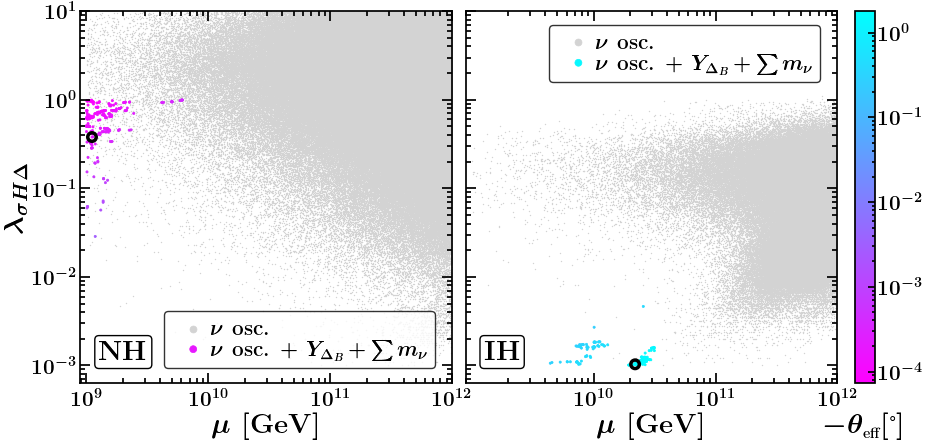}}
\subfloat[]{\label{x1-x2-x3}\includegraphics[scale=0.175,angle=0]{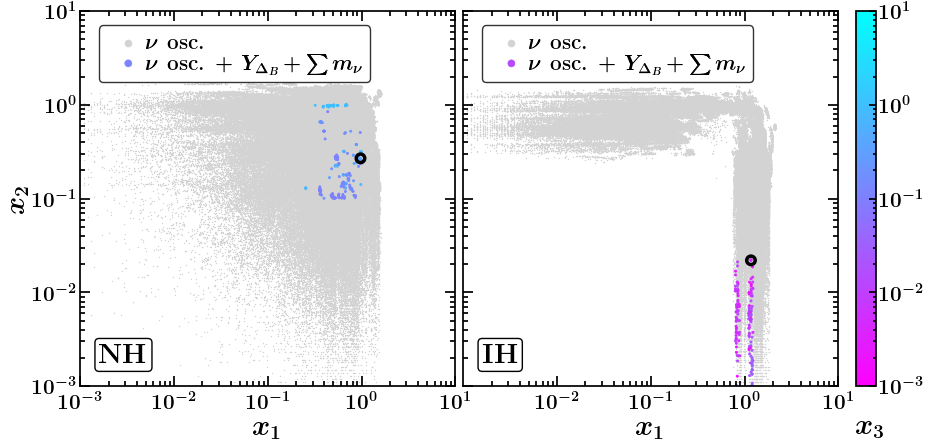}}\\
\hspace*{-1cm}
\subfloat[]{\label{y1-y4-y2}\includegraphics[scale=0.175,angle=0]{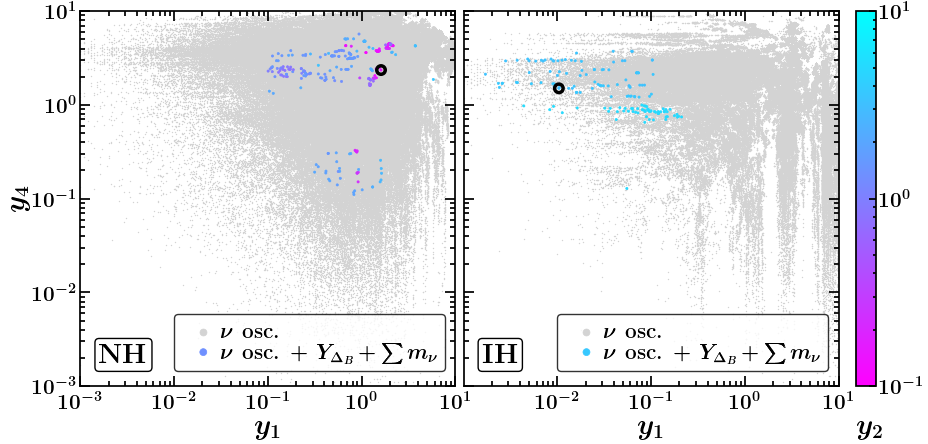}}
\subfloat[]{\label{y3-y5-y6}\includegraphics[scale=0.175,angle=0]{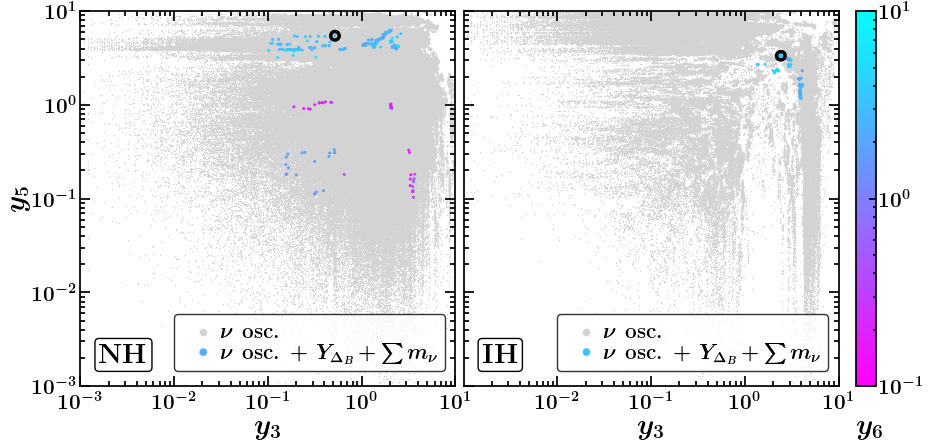}}

\caption{The parameter space which is allowed by neutrino oscillation data within 3$\sigma$ considered in section \ref{sec: nu_osc_data} and additionally is in agreement with the  Planck data for baryon asymmetry within 3$\sigma$ and cosmological upper limit on sum over masses of active neutrinos. The benchmark points given in Table \ref{BP} are denoted by black coloured circle for both hierarchies.
 }
\label{fig:param_space}
\end{center}
\end{figure}

Following the procedure detailed above we are going to estimate the baryon asymmetry generated in our present framework. Our aim is to find the region of parameter space which simultaneously satisfy the neutrino oscillation data \cite{Esteban:2020cvm}, Planck data for baryon asymmetry \cite{Planck:2018vyg} and the Planck data for sum over neutrino masses \cite{Planck:2018vyg}. As a result, in Fig.~\ref{fig:param_space} we show the parameter space which is allowed by neutrino oscillation data considered in section \ref{sec: nu_osc_data} and additionally is in agreement with the Planck data for baryon asymmetry and Planck data for sum over masses of active neutrinos. The black circles present within the two blocks of every panel (for NH and IH) are referred to the two representative benchmark points described in Table \ref{BP}. 

\begin{table}[!h]
\centering
\renewcommand{\arraystretch}{1.5}
\resizebox{15cm}{!}{
\begin{tabular}{|c|c|c|c|c|c|c|c|c|c|c|c|}
\hline 
\multicolumn{12}{|c|}{Model parameters for the benchmark points}\tabularnewline
\hline 
 & $\lambda_{\sigma H \Delta}$ & $\mu$ & $y_1$ & $y_2$ & $y_3$ & $y_4$ & $y_5$ & $y_6$ & $x_1$ & $x_2$ & $x_3$\tabularnewline
\hline 
NH & $0.379$ & $1.110 \times 10^{9}$ & $1.610$ & $0.139$ & $0.519$ & $2.358$ & $5.473$ & $3.106$ & $0.972$ & $0.270$ & $0.409$\tabularnewline
\hline 
IH & $0.001$ & $2.189 \times 10^{10}$ & $0.010$ & $3.228$ & $2.423$ & $1.504$ & $3.349$ & $3.522$ & $1.165$ & $0.022$ & $0.005$\tabularnewline
\hline 
\multicolumn{12}{|c|}{Observables for the benchmark points}\tabularnewline
\hline 
 & $\Delta m_{12}^2$  & $\Delta m_{13}^2$ & $\theta_{12}$ & $\theta_{23}$ & $\theta_{13}$ & $\delta_\text{CP}$ & $J_\text{CP}$ & $\sum m_\nu$ & $Y_B$ & $\epsilon_N$ & $\epsilon_D$\tabularnewline
\hline 
NH & $6.847 \times 10^{-5} \text{ eV}^2$ & $2.444 \times 10^{-3} \text{ eV}^2$ & $31.780 \degree$ & $43.915 \degree$ & $8.804 \degree$ & $359.999 \degree$ & $-1.252 \times 10^{-7}$ & $0.091$ \text{ eV} & $8.754 \times 10^{-11}$ & $-1.546 \times 10^{-7}$ & $-1.321 \times 10^{-9}$\tabularnewline
\hline 
IH & $6.950 \times 10^{-5} \text{ eV}^2$ & $-2.460 \times 10^{-3} \text{ eV}^2$ & $32.286 \degree$ & $48.587 \degree$ & $8.602 \degree$ & $288.068 \degree$ & $-3.110 \times 10^{-2}$ & $0.106$ \text{ eV}  & $8.732 \times 10^{-11}$ & $-1.424 \times 10^{-7}$ & $-1.566 \times 10^{-9}$\tabularnewline
\hline 
\end{tabular}
\renewcommand{\arraystretch}{1.0}
}
\caption{Numerical values for the model parameters and observables for two benchmark points from the allowed parameter space satisfied by all the constraints. The two benchmark points NH and IH represents the normal hierarchy and inverted hierarchy respectively.}
\label{BP}
\end{table}

\begin{figure}[h!]
\begin{center}
\hspace*{-1cm}
\subfloat[]{\label{epN-epD}\includegraphics[scale=0.225,angle=0]{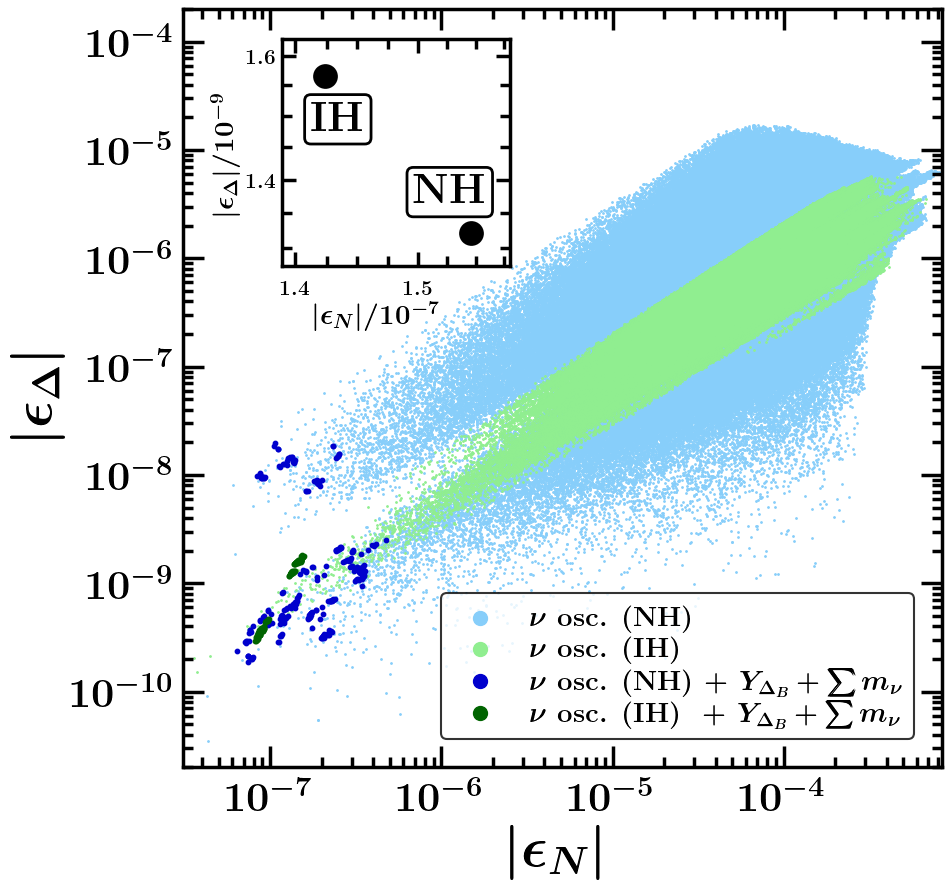}}
\hspace*{0.1cm}
\subfloat[]{\label{YBs}\includegraphics[scale=0.225,angle=0]{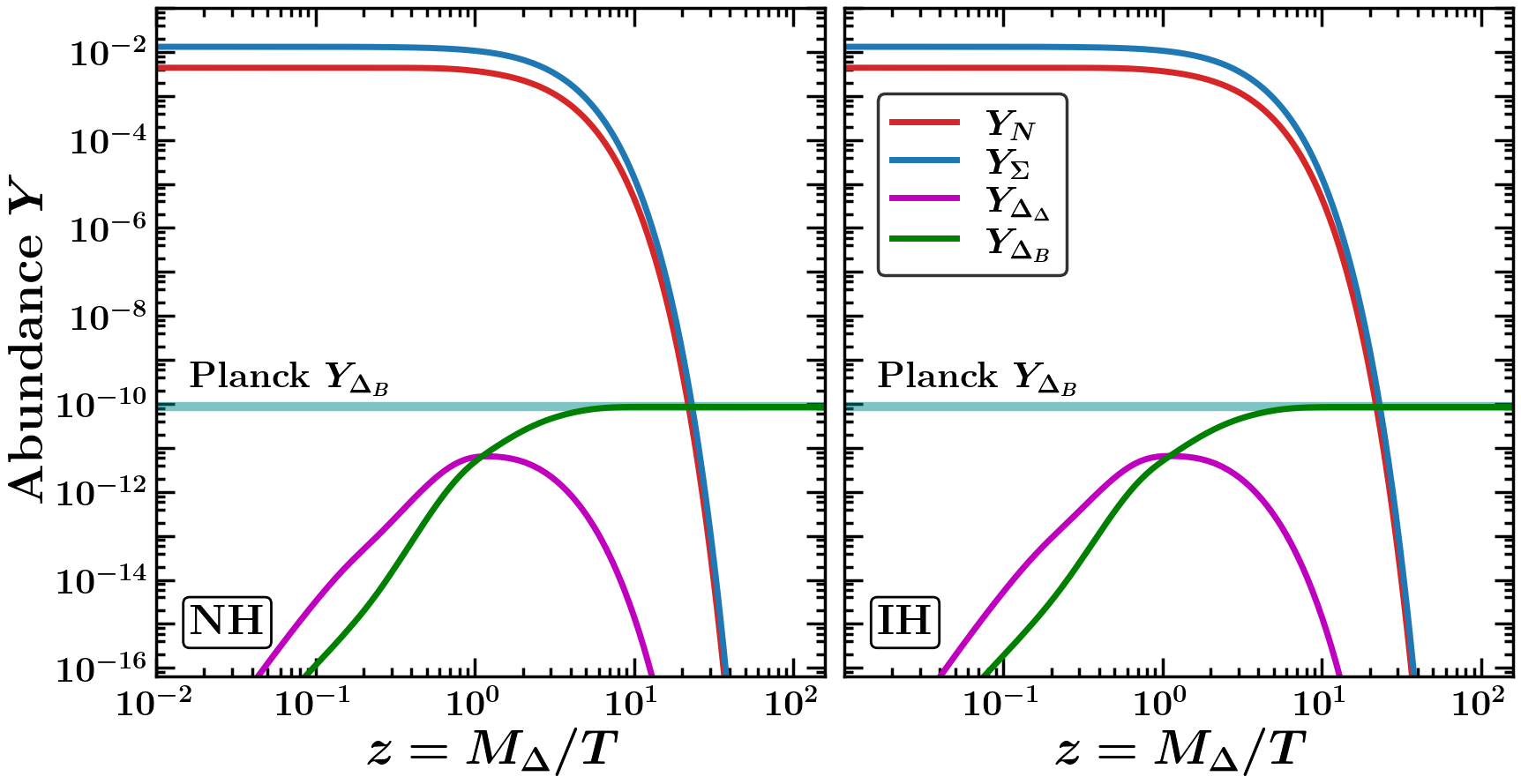}}
\caption{Left panel (\ref{epN-epD}) provides the absolute values of CP asymmetry parameters $\epsilon_{N}$ and $\epsilon_{\Delta}$. In the $|\epsilon_{N}|$-$|\epsilon_{\Delta}|$ plane the region which is covered by small dots represent neutrino oscillation data only. The light blue (green) points represent parameter space that satisfies only neutrino oscillation data in NH (IH). On top of that the region preferred by Planck data for overall baryon asymmetry as well as Planck data for sum over neutrino masses are indicated by large dark blue (green) dots satisfying NH (IH). From the left panel one point for each hierarchy (indicated in the inset by black dots for both NH and IH) has been chosen and for these points the evolution of the comoving number density for each component of lepton asymmetry including the final baryon asymmetry with respect to $z$ have been shown in the two blocks of right panel (\ref{YBs}).}
\label{fig:epsilon_YB}
\end{center}
\end{figure}

In Fig.~\ref{fig:epsilon_YB} we demonstrate the constraints in the parameter space of interest.
The $|\epsilon_{N}|$-$|\epsilon_{\Delta}|$ plane in Fig. \ref{epN-epD} shows the parameter space with green (blue) dots that satisfy the oscillation data for NH (IH). The region preferred by Planck data for overall baryon asymmetry and Planck data for sum over neutrino masses including neutrino oscillation data are indicated by large  dark blue (green) dots for the NH (IH) scenario. From the plot (\ref{epN-epD}) we can estimate the required values of $\epsilon_{N}$ and $\epsilon_{\Delta}$ for which our proposed framework can explain the observed Planck data for BAU. As for example, for the two particular benchmark points (given in the Table \ref{BP}) represented by black dots (as shown in the inset of Fig.~\ref{epN-epD}), for NH (IH) the absolute values of CP asymmetry parameters $\epsilon_{N}$ and $\epsilon_{\Delta}$ are $1.546 \times 10^{-7}$ and $1.321 \times 10^{-9}$ ($1.424 \times 10^{-7}$ and $1.566 \times 10^{-9}$) respectively. Moreover, for these points the evolution of the comoving number density ($Y=n/s$) for each component of lepton asymmetry including the final baryon asymmetry with respect to $z(=M_\Delta/T)$ have been shown in \ref{YBs}. For the purpose of the illustration we take the NH block of the right panel. The red line stands for the evolutions of $N_R$, the blue line shows the evolution of $\Sigma (=\Delta+\Delta^\dagger)$, the magenta line indicates the abundance of $\Delta_\Delta (=\Delta-\Delta^\dagger)$ and the green line represents the evolution of the baryon asymmetry $\Delta_B$ which asymptotically overlaps with the dark cyan horizontal line representing the measured value of BAU at the present epoch. Due to out-of-equilibrium decays of $N_R$ and $\Delta ({\Delta}^\dagger)$ to all possible channels, the red and the blue lines begin to fall around $z=1$. As a consequence the baryon asymmetry increases and around $z=10$ it begins to saturate. In Fig. \ref{YBs} we have assumed an initial thermal abundance for both $N_R$ and $\Delta$. We have checked that relaxing this condition on the initial abundance have minimal numerical impact on the generated lepton asymmetry as has been demonstrated using explicit numerical simulation in Appendix \ref{ap3}.



\section{Conclusion}\label{sec: conclusion}

We present a model of neutrino mass that combine the type I and type II seesaw scenarios in this paper.  An extended scalar sector that includes a singlet and a triplet in addition to the SM Higgs doublet is responsible for breaking CP spontaneously and generating a seesaw mass for the neutrinos.  We utilise a $Z_3$ symmetry to organise the scalar sector that accommodates a complex vev for the singlet scalar. Introduction of a soft  $Z_3$ breaking term necessitated to evade the domain wall problem is also crucial for generating a relative phase between the type I and type II  neutrino mass. This relative phase is the only source of CP violation in this scenario appearing as $\delta_\text{CP}$ in the PMNS matrix while simultaneously driving leptogenesis. 

We set the masses of the scalar triplet and the right handed neutrino at the neutrino mass scale  $\sim 10^{15}$ GeV implying order one real Yukawa couplings. With a single phase generated in the scalar sector we perform an extensive numerical scan using a multi-dimensional Markov Chain Monte Carlo technique to identify the region of parameter space that is in agreement with the neutrino oscillation data.

Next we address the correlated issue of generating matter-antimatter asymmetry through leptogenesis within our framework. Using the single CP phase generated spontaneously in the scalar sector we  demonstrate the possibility of leptogenesis driven by an almost degenerate scalar triplet and right handed neutrino. We scan for the allowed region of parameter space where the novel mixed leptogenesis framework generate the baryon asymmetry while simultaneously satisfying the neutrino oscillation data. 
%
%
\acknowledgments We thank Mainak Chakraborty for discussions. RP acknowledges MHRD, Government of India for the research fellowship. AS acknowledges the financial support from Department of Science and Technology, Government of India through SERB-NPDF scholarship with grant no.:PDF/2020/000245. The authors also acknowledge the National Supercomputing Mission (NSM) for providing computing resources of ‘PARAM Shakti’ at IIT Kharagpur, which is implemented by C-DAC and supported by the Ministry of Electronics and Information Technology (MeitY) and Department of Science and Technology (DST), Government of India.


\begin{appendix}
\renewcommand{\thesection}{\Alph{section}}
\renewcommand{\theequation}{\thesection-\arabic{equation}} 

\setcounter{equation}{0} 

\section{Correlation between PMNS phase $\delta_{\rm CP}$ and scalar CP phase $\theta_\text{eff}$} \label{ap4}
Here we explicitly show that the CP violating phase present in the PMNS matrix is generated from $\theta_{\text{eff}}$ which is the only source of CP violation. Comparing Eq. \ref{eqn: J_in_h} and \ref{eqn: J_in_del} it is straightforward to see that 
\begin{equation}\label{eqn:del_cp_explicit}
\sin\delta_{\rm CP} = \dfrac{8 \Im[h_{12} h_{23} h_{31}]}{\Delta m_{21}^2 \Delta m_{31}^2 \Delta m_{32}^2 \sin(2\theta_{12}) \sin(2\theta_{23}) \sin(2\theta_{13}) \cos\theta_{13}} \ ,
\end{equation}
where $h = M_\nu M_\nu^\dagger$. With $M_\nu$ defined in Eq. \ref{eqn: explicit_nu_mass_matrix} we can express the numerator of Eq. \ref{eqn:del_cp_explicit} in terms of model parameters as given by
\begin{eqnarray}
\Im[h_{12}h_{23}h_{31}] &=& \Im \Bigg[ \frac{e^{-3i\theta_{\rm eff}}}{64 M_R^6}\Bigg(-\sqrt{2}e^{2i\theta_{\rm eff}}M_R v_H^2 v_\Delta x_2(x_1y_1+x_2y_2+x_3y_3)- \nonumber \\
&&\hspace{1em}\sqrt{2}M_R v_H^2 v_\Delta x_1(x_1y_2+x_2y_4+x_3y_5)+e^{i\theta_{\rm eff}}\big(v_H^4x_1x_2(x_1^2+x_2^2+x_3^2) + \nonumber \\
&&\hspace{1em} 2M_R^2v_\Delta^2(y_1y_2+y_2y_4+y_3y_5)\big)\Bigg) \times \nonumber \\
&&\hspace{4.25em} \Bigg(-\sqrt{2}M_R v_H^2 v_\Delta x_3(x_1y_1+x_2y_2+x_3y_3)- \nonumber \\
&&\hspace{1em}  \sqrt{2}e^{2i\theta_{\rm eff}}M_R v_H^2 v_\Delta x_1(x_1y_3+x_2y_5+x_3y_6) + \nonumber \\
&& \hspace{1em} e^{i\theta_{\rm eff}}\left(v_H^4x_1x_3(x_1^2+x_2^2+x_3^2)+2M_R^2v_\Delta^2(y_1y_3+y_2y_5+y_3y_6)\right)\Bigg) \times \nonumber \\
&&\hspace{4.25em} \Bigg(-\sqrt{2}e^{2i\theta_{\rm eff}}M_R v_H^2 v_\Delta x_3(x_1y_2+x_2y_4+x_3y_5)- \nonumber \\
&& \hspace{1em}\sqrt{2}M_R v_H^2 v_\Delta x_2(x_1y_3+x_2y_5+x_3y_6)+e^{i\theta_{\rm eff}}\big(v_H^4x_2x_3(x_1^2+x_2^2+x_3^2) + \nonumber \\
&& \hspace{1em} 2M_R^2v_\Delta^2(y_2y_3+y_4y_5+y_5y_6)\big)\Bigg) \Bigg] \ ,
\end{eqnarray} 
where apart from the exponential factors containing $\theta_{\text{eff}}$, all other parameters are real which ensures that it is the only source of complex phase responsible to generate the CP phase in the PMNS matrix.

\section{Some relevant details for leptogenesis}\label{sec: appendix_a}
In this appendix we present several elements required for leptogenesis.

\subsection{Number Density of  Particle Species} \label{ap1}
Utilising the Maxwell Boltzmann distribution for massive and massless (relativistic) particles, the number densities are given as \cite{AristizabalSierra:2011ab,Sierra:2014tqa}
\begin{eqnarray}
&& n_\Sigma^{eq} (z)=n_\Delta^{eq} (z) +n_{\Delta}^{eq} (z)^\dagger~, \\
&& n_\Delta^{eq} (z)=\frac{3 M_\Delta^3 K_2(z)}{2 \pi^2 z}, \\
&& { n_{N}^{eq} (z) =\frac{M_\Delta^3 r^2 K_2(rz)}{\pi^2 z}}, \\
&& n_{l,H}^{eq} (z)=\frac{2 M_\Delta^3}{\pi^2 z},
\end{eqnarray}
with $r=|M_R|/M_\Delta$ and $K_2(z)$ is the modified Bessel function of second kind. The form of entropy density and Hubble parameter are given below
\begin{eqnarray}
&& s(z)=\frac{4 g^\ast M_\Delta^3}{\pi^2 z^3},\\
&& H(z)=\sqrt{\frac{8 g^\ast}{\pi}}\frac{M^2_\Delta}{M_{\rm Planck} z^2},
\end{eqnarray}
where effective relativistic degrees of freedom $g^\ast=114.5$ and Planck mass $M_{\rm Planck}=1.22\times10^{19}$ GeV.

\subsection{Reaction Densities} \label{ap2}
\begin{figure}[th!]
\begin{center}
\includegraphics[scale=0.275,angle=0]{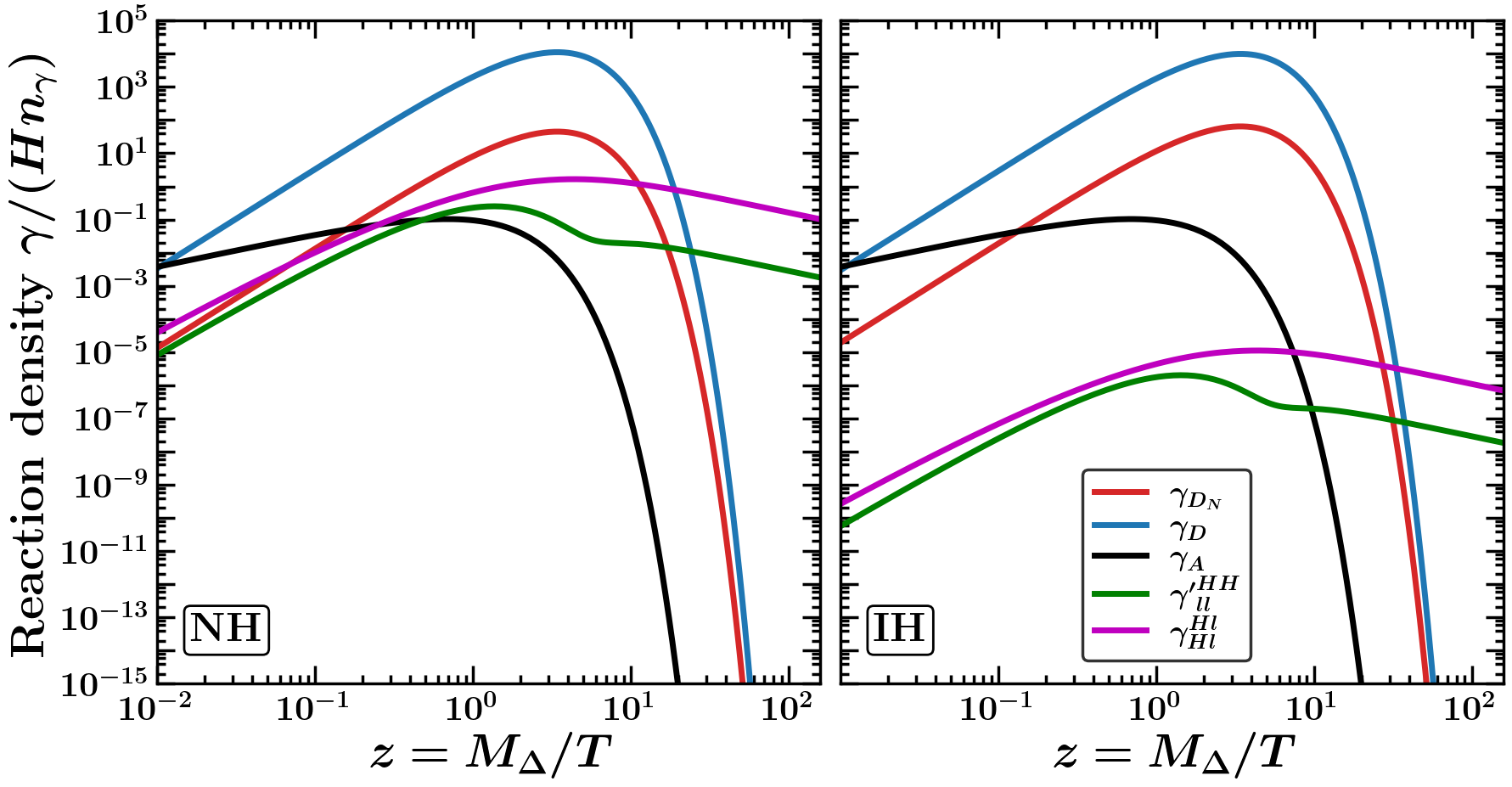}
\caption{Reaction densities (scaled by the product of Hubble parameter and photon number density $\left( n_\gamma (z) = \frac{2}{\pi^2}\frac{M_\Delta^3}{z^3} \right)$ \cite{Pilaftsis:2003gt}) with respect to $z$ for different processes involved in the present hybrid unflavoured leptogenesis model. The left (right) panel is for NH (IH). The curves are depicted for the two benchmark points (indicated in the inset of Fig. \ref{epN-epD}) given in the Table \ref{BP}.}
\label{fig:reaction_density}
\end{center}
\end{figure}
Decay $(1\rightarrow 2)$ related reaction densities for $N_R$ and $\Delta$ are \cite{AristizabalSierra:2011ab,Sierra:2014tqa}
\begin{eqnarray}
&& { \gamma_{D_{N}}=\frac{1}{8 \pi^3} \frac{M_\Delta^5 r^4 K_1(rz) (\mathcal{Y}_\nu \mathcal{Y}_\nu^\dagger)_{11} }{z |M_{R}| }}; \quad \gamma_D= \frac{K_1(z)}{K_2(z)} n_\Sigma^{eq} (z) \Gamma_\Delta\;.
\end{eqnarray}
The general form of $(2\leftrightarrow2)$ scattering reaction densities is

\begin{equation}
 \gamma_s= \frac{M_\Delta^4}{64 \pi^4} \int_{x_{min}}^{\infty}  \sqrt{x} \frac{K_1(z\sqrt{x}) \widehat{\sigma}_s}{z} dx \label{gen_s}\;,
\end{equation}
with $x=\tilde{s}/M_{\Delta}^2$ ($\tilde{s}$ centre of mass energy) and $\widehat{\sigma}_s$ is referred as the reduced cross section.
For Yukawa induced process $x_{min}=0$ and gauge induced process it is $x_{min}=4$. The reduced cross sections for the gauge mediated processes is \cite{AristizabalSierra:2011ab,Sierra:2014tqa}
\begin{eqnarray}
 \widehat{\sigma}_A & = &\frac{2}{72 \pi} \Big \{ (15 C_1-3 C_2) \omega + (5 C_2 -11 C_1 ) \omega^3 + 3 (\omega^2 -1)[2 C_1 + C_2 (\omega^2 -1) ] \ln \Big (\frac{1+\omega}{1- \omega} \Big) 
 \Big \} \nonumber \\ & + &
 \Big  (  \frac{50 g_{2L}^4 +41 {g_{1Y}}^4}{48 \pi} \Big ) \omega^{\frac{3}{2}}, \label{gauge_s}
\end{eqnarray}
with $\omega \equiv \omega (x) =\sqrt{1-4/x}$ and $C_1= 12 g_{2L}^4 +3 g_{1Y}^4 +12 g_{2L}^2 g_{1Y}^2,~C_2=6 g_{2L}^4+3 g_{1Y}^4 + 12 g_{2L}^2 g_{1Y}^2$. 

In Fig.\;\ref{fig:reaction_density} we have shown the reaction densities for the several processes for the benchmark points as given in Table \ref{BP} for both NH and IH.

\begin{figure}[htbp!]
\begin{center}
\includegraphics[scale=0.275,angle=0]{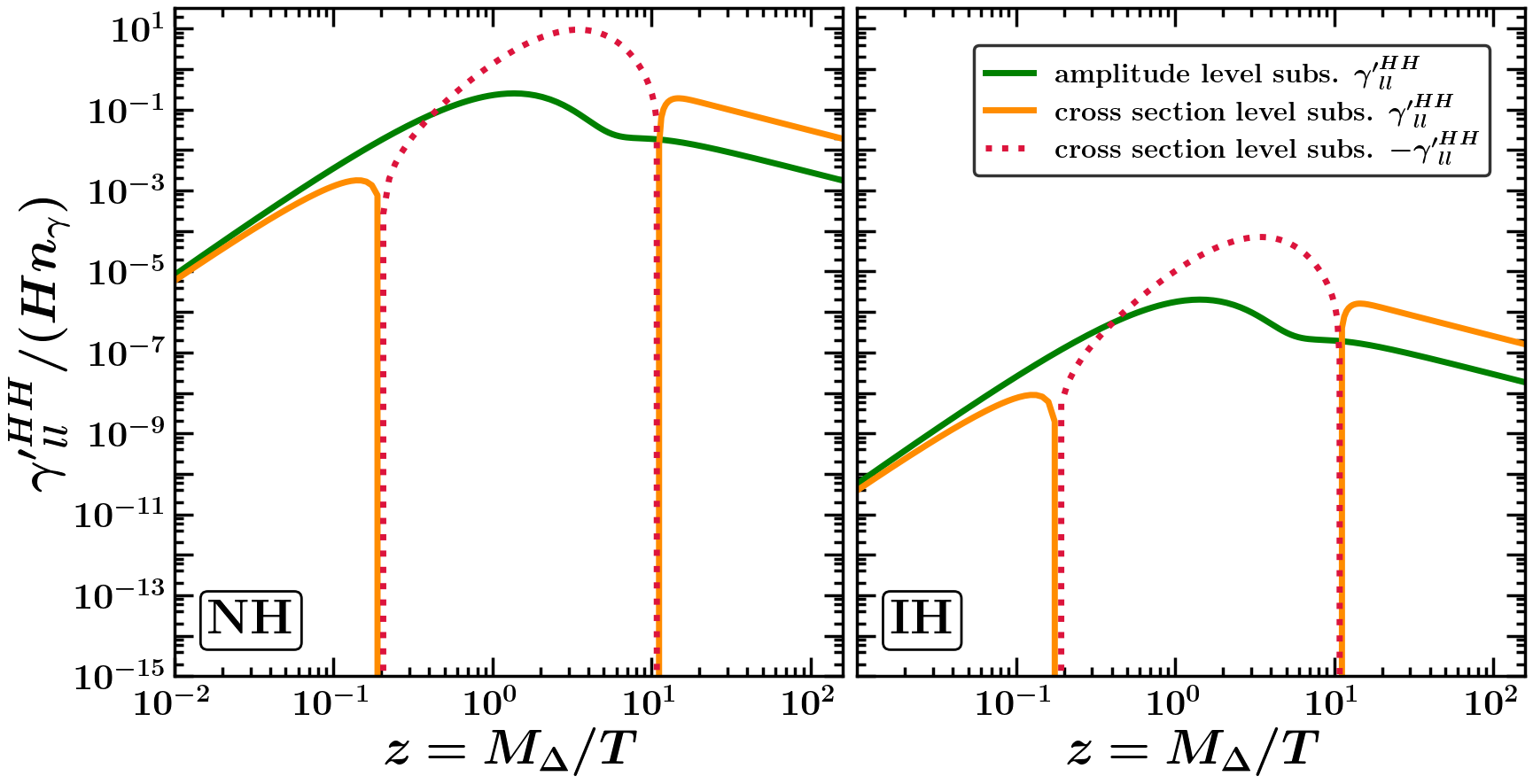}
\caption{Reaction density (scaled by the product of Hubble parameter and photon number density) with respect to $z$ for the process $HH \to ll$ for both NH and IH for both the schemes of RIS. The left (right) panel is for NH (IH). The curves are depicted for the two benchmark points (indicated in the inset of Fig. \ref{epN-epD}) given in the Table \ref{BP}. We have used the subtraction scheme indicated by the green solid line in our work.}
\label{fig:compare_ris}
\end{center}
\end{figure}
A few comments about the real intermediate state (RIS) subtraction scheme which is necessary to avoid double counting for $2\to 2$ scattering processes are now in order. Various approaches have been taken to implement RIS subtraction mechanism in the literature \cite{Giudice:2003jh, Pilaftsis:2003gt}. To avoid negative values we have adopted the subtraction method at the amplitude level by taking into account only the principal part of the propagator in the transition amplitude as has been advocated in \cite{Luty:1992un, Plumacher:1996kc, Buchmuller:2002jk, Buchmuller:2003gz}. This may be contrasted with the subtraction scheme at the level of cross section \cite{Sierra:2014tqa} as can be seen in Fig. \ref{fig:compare_ris}.
\subsection{Comparison between initial abundances} \label{ap3}
\begin{figure}[ht!]
\centering
\includegraphics[scale=0.275,angle=0]{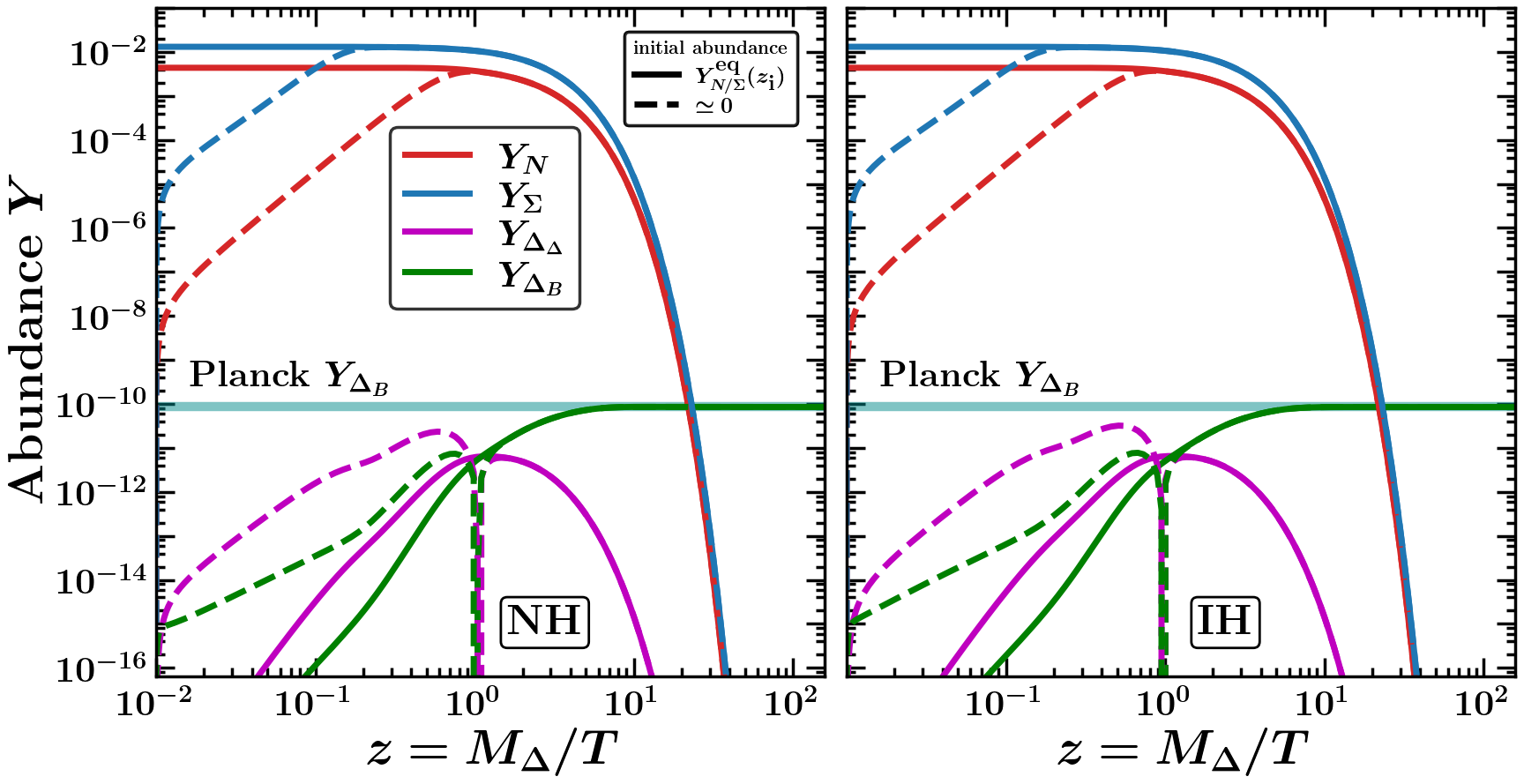}
\caption{Evolution of various asymmetries considering $N_R$ and $\Delta$ following initial thermal equilibrium number density (solid lines) and vanishing initial abundance (dashed lines) for the benchmark points of NH (left panel) and IH (right panel) given in Table \ref{BP}. }
\label{fig: zero_init_abundance}
\end{figure}
Here we illustrate the comparison between the evolution of the abundances of various particle species along with the final baryon asymmetry if one considers vanishing initial abundance of RHN and the triplet with that of their equilibrium number densities. In case of vanishing initial abundance of $N_R$ and $\Delta$, the red and blue lines start to rise upto a value of $z = z^\text{eq}$ given by $Y_{N/\Delta}(z^\text{eq}) = Y_{N/\Delta}^\text{eq}(z^\text{eq})$ which is in agreement with \cite{Buchmuller:2004nz}. The nature of evolution of the baryon asymmetry is different in these two cases, however, the final value remains same as can be interpreted from Fig. \ref{fig: zero_init_abundance}.

\end{appendix}

\newpage
\bibliographystyle{JHEP}
\bibliography{sdt_z3}
\end{document}